# The Hα Luminosity Function of Galaxies at $z \sim 4.5$

Victoria Bollo[1,2], Valentino González[1], Mauro Stefanon[3,4,5], Pascal A. Oesch[6,7], Rychard J. Bouwens[3],
Renske Smit[8], Garth D. Illingworth[9], and Ivo Labbé[10]
[1] Departamento de Astronomía, Universidad de Chile, Camino del Observatorio 1515, Las Condes, Santiago 7591245, Chile; vbollo@das.uchile.cl
[2] European Southern Observatory (ESO), Karl-Schwarzschild-Strasse 2, D-85748 Garching bei München, Germany
[3] Leiden Observatory, Leiden University, NL-2300 RA Leiden, Netherlands
[4] Departament d'Astronomia i Astrofísica, Universitat de València, C. Dr. Moliner 50, E-46100 Burjassot, València, Spain
[5] Unidad Asociada CSIC "Grupo de Astrofísica Extragaláctica y Cosmología" (Instituto de Física de Cantabria - Universitat de Valencia), Spain
[6] Departement d'Astronomie, Universitè de Genève, 51 Ch. des Maillettes, CH-1290 Versoix, Switzerland
[7] Cosmic Dawn Center (DAWN), Niels Bohr Institute, University of Copenhagen, Jagtvej 128, KØbenhavn N, DK-2200, Denmark
[8] Astrophysics Research Institute, Liverpool John Moores University, 146 Brownlow Hill, Liverpool L3 5RF, UK
[9] UCO/Lick Observatory, University of California, Santa Cruz, 1156 High Street, Santa Cruz, CA 95064, USA
[10] Centre for Astrophysics and SuperComputing, Swinburne, University of Technology, Hawthorn, Victoria, 3122, Australia
Received 2022 June 6; revised 2023 February 13; accepted 2023 February 14; published 2023 April 7

## Abstract

We present the Hα luminosity function (LF) derived from a large sample of Lyman break galaxies at $z \sim 4.5$ over the GOODS-South and North fields. This study makes use of the new, full-depth Spitzer/IRAC [3.6] and [4.5] imaging from the GOODS Re-ionization Era wide-Area Treasury from the Spitzer program. The Hα flux is derived from the offset between the continuum flux estimated from the best-fit spectral energy distribution, and the observed photometry in IRAC [3.6]. From these measurements, we build the Hα LF and study its evolution providing the best constraints of this property at high redshift, where spectroscopy of Hα is not yet available. Schechter parameterizations of the Hα LF show a decreasing evolution of $\Phi^*$ with redshift, increasing evolution in $L^*$, and no significant evolution in the faint-end slope at high $z$. We find that star formation rates (SFRs) derived from Hα are higher than those derived from the rest-frame UV for low SFR galaxies but the opposite happens for the highest SFRs. This can be explained by lower mass galaxies (also lower SFR) having, on average, rising star formation histories (SFHs), while at the highest masses the SFHs may be declining. The SFR function is steeper, and because of the excess SFR(Hα) compared to SFR(UV) at low SFRs, the SFR density estimated from Hα is higher than the previous estimates based on UV luminosities.

*Unified Astronomy Thesaurus concepts:* Galaxy evolution (594); Galaxy formation (595); High-redshift galaxies (734)

## 1. Introduction

A key aspect in understanding the formation and evolution of galaxies across cosmic time is the study of their star formation rate (SFR). While multiwavelength galaxy surveys have played a key role in identifying large numbers of galaxies all the way to $z \sim 11$ (Eyles et al. 2005; Vanzella et al. 2005, 2006, 2008, 2009; Verma et al. 2007; Stark et al. 2009; Yabe et al. 2009; González et al. 2010, 2012; Labbé et al. 2010; Grogin et al. 2011; Kashikawa et al. 2011; Koekemoer et al. 2011; Trenti et al. 2011; Windhorst et al. 2011; Brammer et al. 2012; Finkelstein et al. 2012; Bouwens et al. 2014, 2015; Skelton et al. 2014; Kriek et al. 2015; Oesch et al. 2015, 2018; van Dokkum et al. 2013; Hasinger et al. 2018; Ono et al. 2018; Jiang et al. 2021), estimating SFRs consistently at all redshifts has proven to be very challenging (Katsianis et al. 2017a, 2017b).

Hα is one of the most used estimators of the SFR of galaxies up to $z \lesssim 3$ (Erb et al. 2006; Hanish et al. 2006; Geach et al. 2008; Förster Schreiber et al. 2009; Hayes et al. 2010; Reddy et al. 2010; Ly et al. 2011; Weisz et al. 2012; Sobral et al. 2013, 2016; Stroe et al. 2017; Coughlin et al. 2018). As an SFR estimator, it depends on the ionizing flux from the most massive stars (with lifetimes <10 Myr), it is not sensitive to the metallicity of the gas (although the ionizing flux of stars depend on the stellar metallicity), and because of its wavelength, it is less affected by dust obscuration compared to estimators at shorter wavelengths. By comparison, the most commonly used SFR estimator at high-$z$ ($z > 3$), the rest-frame UV luminosity traces the light of slightly lower mass stars (lifetimes ∼100 Myr) and can be up to 3× more affected by dust obscuration. Because of the different timescales, the comparison between SFR estimates derived from Hα (SFR$_{H\alpha}$) and from rest-frame UV luminosity (SFR$_{UV}$) could be informative about the star formation histories (SFHs) of galaxies and represents a major challenge (Kennicutt & Evans 2012; Madau & Dickinson 2014; Smit et al. 2016; Emami et al. 2019). For example, Atek et al. (2022) show that low-mass galaxies at $0.7 < z < 1.5$ tend to have an elevated SFR(Hα) compared to SFR(UV), which they interpret as short-time variations in the SFR, or *burstiness*. Similar differences have also been found in other studies, as reported by Katsianis et al. (2017a, 2017b), suggesting that they may be more prominent at higher redshifts. Shivaei et al. (2015), however, investigated the SFR$_{H\alpha}$ and SFR$_{UV}$ at $z \sim 2$ and concluded that they are roughly consistent when using a Calzetti et al. (2000) attenuation curve (with the same normalization for the stellar continuum and the nebular emission).

Unfortunately, Hα is not readily observable at redshifts $z > 2.8$, when it shifts to wavelengths beyond the $K$ band (the situation will soon change dramatically thanks to JWST). For this reason, most SFR studies at $z > 3$ tend to be based exclusively on the rest-frame UV luminosity. There is, however, a redshift window that can be exploited to estimate the H$\alpha$ flux at high redshift, in particular, using deep Spitzer/IRAC photometry. The idea is that at specific redshifts, strong nebular lines in the rest-frame optical contribute to the flux measured in one of the IRAC bands, while others sample strictly the stellar continuum. This color offset can be used to estimate the flux of the nebular lines. Several studies have taken advantage of this offset to infer the intensity of nebular emission lines at $z > 3$ (Shim et al. 2011; Stark et al. 2013; Shivaei et al. 2015; Marmol-Queralto et al. 2016; Rasappu et al. 2016; Smit et al. 2016; De Barros et al. 2019; Caputi et al. 2017; Faisst et al. 2017), and even at $z \sim 8$ (Stefanon et al. 2022).

The ability to estimate the contribution of nebular emission lines to Spitzer/IRAC photometry depends strongly on the depth of the IRAC imaging. In this work, we take advantage of the new, full-depth Spitzer/IRAC [3.6] and [4.5] imaging from the GOODS Re-ionization Era wide-Area Treasury from Spitzer (GREATS) program (Stefanon et al. 2021), reaching up to 250 hr of integration. We exploit the redshift window in which we can isolate the contribution of the H$\alpha$ line to estimate the H$\alpha$ luminosity function (LF) for the first time at $z \sim 4.5$. We explore standard corrections for dust attenuation to estimate intrinsic selection, which makes use of ultradeep, wide luminosities, and derive the SFR(H$\alpha$), the SFR function and its integral, the cosmic star formation rate density (CSFRD) at $z \sim 4.5$.

This paper is organized as follows. In Section 2, we describe the data that have been used and in Section 3 how the final sample of spectroscopic and photometric redshift galaxies was selected. Section 4 is concerned with the methodology used to measure the H$\alpha$ flux, describing the method used to derive it and its limitations. In Section 5, we derive the H$\alpha$ LF and its best-fit Schechter parameterization. In Section 6, we derive SFRs from the H$\alpha$ fluxes and from the UV luminosities and compare them. We also derive the SFR function at $z \sim 4.5$. In Section 7, we discuss our findings, and compare them with previous studies and other SFR tracers commonly used at high redshift. A summary of the main results is presented in Section 8. Throughout this paper, we use $H_0 = 70$ km s$^{-1}$ Mpc$^{-1}$, $\Omega_m = 0.3$, and $\Omega_\Lambda = 0.7$. Magnitudes are quoted in the AB systems (Oke & Gunn 1983).

## 2. Data

### 2.1. Sample Selection and HST Data

This work is based on the Lyman break galaxy (LBG) selection by Bouwens et al. (2015), focusing in particular on the sources at $z \sim 4$ and $z \sim 5$ (B- and V-band dropouts, respectively) found over the GOODS fields (Giavalisco 2002). Their selection makes use of ultradeep, wide-area observations obtained as part of the CANDELS program over the GOODS-North, GOODS-South fields, the ERS field (Windhorst et al. 2011), and the UDF/XDF (Beckwith et al. 2006; Illingworth et al. 2013) field. The available photometry from the Hubble Space Telescope (HST) includes the $B_{435}$, $V_{606}$, $i_{775}$, $I_{814}$, $z_{850}$, $J_{125}$, $JH_{140}$, and $H_{160}$ bands, reaching 5$\sigma$ depths between 26.2 and 28 in the CANDELS fields, between 26.4 and 27.7 in the ERS field, and ranging from 29.2–30 in the XDF field. The total search area corresponds to $\sim$300 arcmin$^2$ where Bouwens et al. (2015) identified 7574 star-forming galaxy candidates at $z \geqslant 3$, selected as B- or V-band dropouts.

The samples at $z \sim 4$ and $z \sim 5$ were selected by Bouwens et al. (2015) (see Figure 1 of their paper) using the following LBG criteria:

1. At $z \sim 4$:

$$(B_{435} - V_{606} > 1) \wedge (i_{775} - J_{125} < 1) \wedge$$
$$(B_{435} - V_{606} > 1.6(i_{775} - J_{125}) + 1) \wedge$$
$$(\text{not in } z \sim 5 \text{ selection})$$

2. At $z \sim 5$:

$$(V_{606} - i_{775} > 1.2) \wedge (z_{850} - H_{160} < 1.3) \wedge$$
$$(V_{606} - i_{775} > 0.8(z_{850} - H_{160}) + 1.2) \wedge$$
$$(S/N(B_{435}) < 2) \wedge (\text{not in } z \sim 6 \text{ selection}).$$

The most significant source of contamination in the sample are lower redshift galaxies that spuriously satisfy the color–color criteria due to the effect of photometric noise (see Section 3.5.5 in Bouwens et al. 2015). However, this represents minimal contamination since it was carefully estimated by adding noise to real observations, providing a direct and robust estimate. Overall, the contamination rates produced by stars, transient sources, lower redshift objects, extreme emission line galaxies, and spurious sources were estimated to be a total level of contamination of just $\sim$2% and $\sim$3% for the $z \sim 4$ and $z \sim 5$ samples, respectively.

Our initial sample contains 5712 B-band dropouts at $z \sim 4$ and 1862 V-band dropouts at $z \sim 5$.

### 2.2. GREATS Spitzer/IRAC Photometry

In this work, we will measure H$\alpha$ fluxes based on the impact that the line has on broadband photometry. At $z > 3.8$, this requires Spitzer/IRAC imaging at 3.6 and 4.5 $\mu$m. Here we take advantage of new full-depth Spitzer/IRAC 3.6 $\mu$m and 4.5 $\mu$m imaging from the GOODS Re-ionization Era wide-Area Treasury from Spitzer (GREATS) program (PI: I. Labbé Stefanon et al. 2021) over the GOODS-N and GOODS-S fields.

The GREATS data set extends the ultradeep coverage in the [3.6] and [4.5] bands with >150 hr of deep data (corresponding to a 1$\sigma$ sensitivity of 28.7 and 28.3 in the [3.6] and [4.5] bands) across $\sim$150 arcmin$^2$ ($\sim$1/2 total area of the GOODS fields). The GREATS mosaics reach an impressive 250 hr coverage in a small $\sim$5–10 arcmin$^2$ region in each field in the [3.6] and [4.5] bands. The available coverage in the [5.8] and [8.0] bands is shallower. In the GOODS-N field, the maximum coverage is $\sim$90 hr, corresponding to a 1$\sigma$ depth of 26.0 and 25.8 for the [5.8] and [8.0] bands. For the GOODS-S field, the maximum depth is $\sim$40 hr, corresponding to 1$\sigma$ limits of 25.6 and 25.4 in the [5.8] and [8.0] bands.

The deep imaging and slightly low resolution of the [3.6] and [4.5] mosaics create source blending issues that may limit our ability to perform photometry (the confusion limit). Here we make use of MOPHONGO (Labbé et al. 2015), a source deblending software that exploits the high-resolution imaging available from HST on the same fields to model the light profile of all sources in the field and remove possible contamination from nearby sources.

To test the performance of the code on our data, Stefanon et al. (2021) performed Monte Carlo (MC) simulations consisting of injecting synthetic point sources at random positions. Then, their flux densities were measured with MOPHONGO and corrected to total using the brightness profile of each source on the low-resolution image and the point spread function reconstructed at the specific locations of each source. They showed that the code recovers the fluxes of the synthetic sources within the expected noise independent of luminosity, with only a small fraction ($\lesssim$10%) of sources deviating appreciably ($>5\sigma$) from the true flux. Therefore, source confusion in deep IRAC imaging can be reliably mitigated even in the faintest regimes (see Stefanon et al. 2021 for details).

## 3. A Sample for H$\alpha$ Measurements at $z \sim 4.5$

Between redshift 3.86 and 4.94, the H$\alpha$ line can contribute to the flux measured with Spitzer/IRAC in the [3.6] $\mu$m band. Starting from the original sample of $B$- and $V$-band dropouts, we have imposed restrictions on the quality of the IRAC photometry and the redshifts to estimate H$\alpha$ on a reliable subsample.

### 3.1. IRAC Photometry

As described above, MOPHONGO mitigates the problem of source confusion in deep IRAC images. The code automatically flags poor neighbor subtraction but we have also chosen to inspect by eye all the residual images, discarding the sources with strong residuals in the area where we perform aperture photometry. Since this criterion depends primarily on the ability to model the neighboring sources, it is not expected to introduce any significant biases in the sample.

As will be discussed later (see Section 4), our method to estimate H$\alpha$ from broadband photometry relies on accurately estimating the stellar continuum flux in the rest-frame optical, near the wavelength of H$\alpha$. This requires an appropriate signal-to-noise ratio (S/N) in both IRAC bands. We have imposed an S/N > 2 in both IRAC bands simultaneously but since [4.5] is shallower than [3.6], this is primarily a limit on [4.5] luminosity. Since [4.5] does not include a contribution from emission lines, this is also approximately a limit in stellar mass.

### 3.2. Redshifts

At redshifts $3.86 < z < 4.94$, the H$\alpha$ line contributes flux to the IRAC [3.6] band (e.g., see Figure 1 in Smit et al. 2016, and Figure 3 in this paper). Accurate redshifts are important to ensure that H$\alpha$ is at the right wavelength so our work makes use of spectroscopic redshifts when available but it also uses photometric redshifts for which the line is at the right wavelength with high confidence.

#### 3.2.1. Spectroscopic Sample

We match our $B$- and $V$-band dropout catalogs with the spectroscopic samples by Herenz et al. (2017), Oyarzún et al. (2017, 2016), Shim et al. (2011), Stark et al. (2013), Vanzella et al. (2005, 2006, 2008, 2009), and Balestra et al. (2010) over the GOODS-South and GOODS-North fields. Redshifts in these works are mainly derived from prominent features such as Ly$\alpha$ in emission but there are also redshifts based on UV absorption features. As a result, we have found 69 $B$-band dropouts and 21 $V$-band dropouts with spectroscopic redshift between $z = 3.96$ and 4.94.

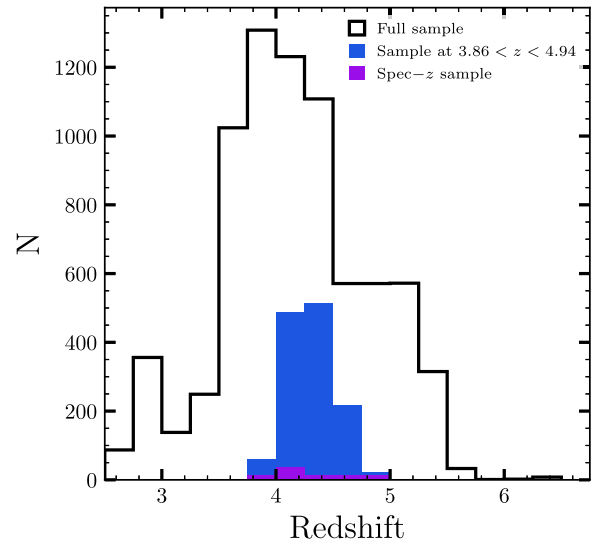

**Figure 1.** Comparison between the redshift selected samples at $3.86 < z < 4.94$ (blue histogram) and the parent sample of $B$- and $V$-band dropouts (black histogram). The selected sample includes $B$- and $V$-band dropouts, according to their lack of brightness in the $B_{435}$ and $V_{606}$ bands, respectively, and the redshifted H$\alpha$ emission falls in the measurement range of the 3.6 $\mu$m band. The median redshift of the sample is $z \sim 4.3$, taking into account the $z$-spec and $z$-phot samples. The selection criteria that assess the quality of data are the same in both samples (see Section 3) regarding their reliable IRAC photometry (e.g., cuts based on IRAC S/N, quality of the SED fit) and accurate redshifts.

#### 3.2.2. Photometric Redshift Sample

As will be discussed in the next section, our fiducial estimates of H$\alpha$ fluxes make use of spectral energy distribution (SED) fitting, which in the case of sources without spectroscopic redshift, includes fitting for the best redshift. For the SED fitting we use the code CIGALE (Boquien et al. 2019; more details in Section 4.1), which can estimate the probability distribution function (PDF) for the redshift (marginalizing over all other parameters of the fit while assuming a flat prior). We model the photometric redshift excluding the IRAC bands with possible nebular emission contamination. We select sources that have at least an 80% probability of being at the desired redshift range. This results in 1299 sources at $z = 3.86-4.94$, for which H$\alpha$ contributes flux to the [3.6] band.

#### 3.2.3. Spectroscopic Sample versus Photometric Sample

Figure 1 shows the redshift distribution of the parent sample of $B$- and $V$-band dropouts as well as the spectroscopic and photometric redshift samples. Figure 2 further compares the spectroscopic and photometric redshift samples at $z \sim 4.5$. The differences in redshift distribution are most likely explained by the selection function and the sizes of the parent samples of the different studies included in the spectroscopic sample. The UV-continuum slope, $\beta$ was determined fitting all the fluxes between the Lyman break and the Balmer break ($z_{850}$, $J_{125}$, $JH_{140}$, and $H_{160}$) with a power law $f_\lambda \propto \lambda^\beta$. The left panel shows that both samples have very similar rest-UV colors, with median $\beta$ values of $-1.92$ and $-1.94$ for the spectroscopic and photometric samples, respectively. The right panel shows the distribution of the UV-to-optical color, $H_{160} - [4.5]$, indicating slightly bluer colors in the photo-$z$ sample, probably indicative of slightly younger ages.

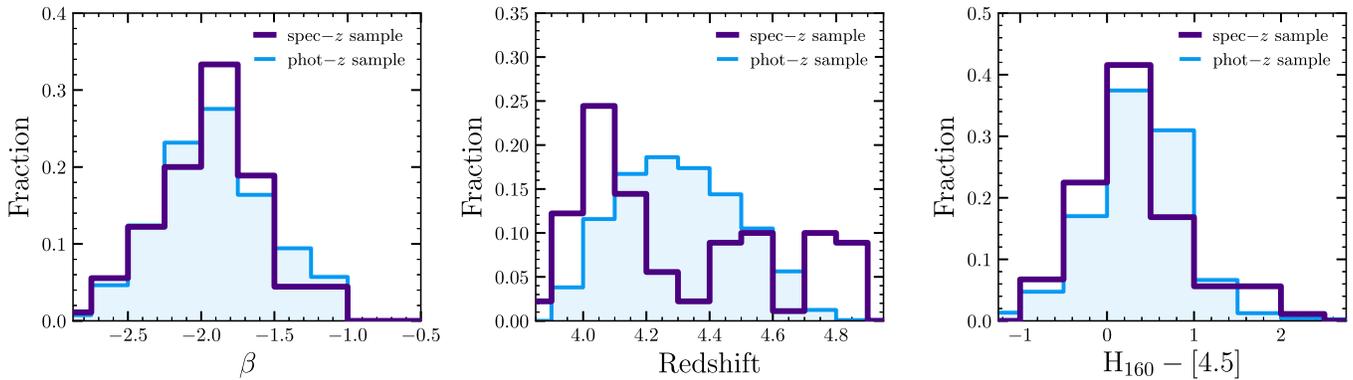

**Figure 2.** Comparison of the observational properties of our photo-$z$ sample (blue-filled histograms) and our spec-$z$ sample (purple histograms) for the redshift selected sample at $z \sim 4.5$. Left panel: the UV-continuum slope $\beta$, defined as $f_\lambda \propto \lambda^\beta$, shows a similar distribution in both samples, with a median of $-1.94$ for the photo-$z$ sample, and $-1.92$ for the spec-$z$ sample. Middle panel: the redshift distribution of the two samples shows differences for both samples, the median value of the spec-$z$ sample of 4.15 is slightly less than the median value 4.30 of the photo-$z$ sample. Right panel: the $H_{160} - [4.5]$ color, taking advantage of the rest-frame UV continuum given by $H_{160}$ and the rest-frame optical continuum flux of the 4.5 $\mu$m band. The median for the photo-$z$ sample is 0.36, and for the spec-$z$ sample is 0.25. Despite these differences, there does not seem to be any significant bias between both the photometric and spectroscopic samples at $z \sim 4.5$.

## 4. H$\alpha$ Measurements

The key to measuring the H$\alpha$ flux through broadband photometry is to estimate independently the level of the underlying continuum. In the sections below we will focus on the standard method that uses the excess between the best-fit SED and the flux measured from the photometry (Shim et al. 2011; Stark et al. 2013; Shivaei et al. 2015; Marmol-Queralto et al. 2016; Rasappu et al. 2016; Smit et al. 2016). As a check, we have also applied an alternative method that interpolates the underlying continuum based on a sample of sources at a slightly lower redshift ($3.0 < z < 3.7$) for which IRAC colors are unaffected by line emission. This alternative method, which is fully empirically based and is independent of the choice of stellar population models, produces very consistent results. For more details on this alternative method, see Appendix A.

### 4.1. SED Modeling

To estimate the stellar continuum at the wavelength of H$\alpha$, we use the code CIGALE (Boquien et al. 2019) to fit synthetic stellar population models to the observed rest-frame UV-to-optical photometry. We exclude photometry with a possible contribution from nebular emission to ensure that what we estimate is the underlying stellar continuum only. Thus, we excluded the observed photometry at [3.6] $\mu$m for the sample selected in the redshift range of $z = 3.86 - 4.94$, where the redshifted H$\alpha$ emission line boosts the observed flux.

We performed SED fitting using fairly flexible SFHs to accurately interpolate the stellar continuum flux at the wavelength of H$\alpha$. Briefly, our models use the simple stellar population models from Bruzual & Charlot (2003) with a Chabrier (2003) initial mass function (IMF) and we fixed the metallicity at $0.2 Z_\odot$. The SFHs were set to a double exponential with $e$-folding times of $\tau_{\rm main} = 150$ Gyr (essentially a constant SFH), and $\tau_{\rm burst} = 10$ Myr. The relative mass fraction of the late burst to the older population is allowed to vary between 0 and 0.95, and the ages of each exponential SFH are varied between 30 Myr and the age of the universe at the lowest redshift in the grid. This allows us to simultaneously fit fairly evolved (old) stellar populations as well as recent bursts. For internal reddening, we used the Calzetti et al. (2000) attenuation curve, with the dust extinction allowed to vary between $E(B - V) = 0$ and $E(B - V) = 0.6$. To estimate the uncertainty of the flux at the wavelength of H$\alpha$, we run a set of 100 realizations, in which the input photometric measurements are perturbed according to their uncertainties.

Finally, the redshift grid is allowed to vary between $z = [2.5 - 5.5]$ for $B$-band dropouts and between $z = [3.5 - 6.5]$ for $V$-band dropouts. From these, we selected galaxies in the redshift window of $3.86 < z_{\rm phot} < 4.94$ (the $z \sim 4.5$ sample) for which the redshifted H$\alpha$ falls in the measurement range of the IRAC [3.6] band. Figure 1 shows a histogram of the redshifts for the $z \sim 4.5$ sample (blue). Figure 3 shows two SED examples with their respective best fits for a spec-$z$ sample (top) and a photo-$z$ sample (bottom). As explained above, open circles are ignored in the SED fitting process to avoid nebular lines influencing our pure stellar fits.

CIGALE can also produce models that include the nebular emission (both continuum and lines, e.g., Stark et al. 2013) so we also produced SED fits that use all the photometry and fits synthetic models with nebular emission. These models can directly output the best-fit H$\alpha$ flux for each galaxy. We consider an ionization parameter of $\log U = [-3.5, -2.0]$ in steps of 0.1, and all other parameters in the grid of models are identical as in the previous fits. We find that the H$\alpha$ fluxes estimated this way are comparable to the more standard method adopted in the rest of the paper (but slightly biased to higher values, for details see Appendix A.1).

### 4.2. H$\alpha$ Flux Measurements

To estimate the H$\alpha$ flux we compare the observed broadband photometry, which includes nebular emission, with that expected from the best-fit models produced by CIGALE using templates without nebular emission. For the sample at $z \sim 4.5$, IRAC [3.6] is the relevant passband. To determine the H$\alpha$ flux we compare the observed photometry with the synthetic photometry from the models finding a systematic excess in the samples as shown in Figure 3. We determine the flux of a single Gaussian at the wavelength of H$\alpha$ that would reproduce the observed excess. In practice, the excess could also be

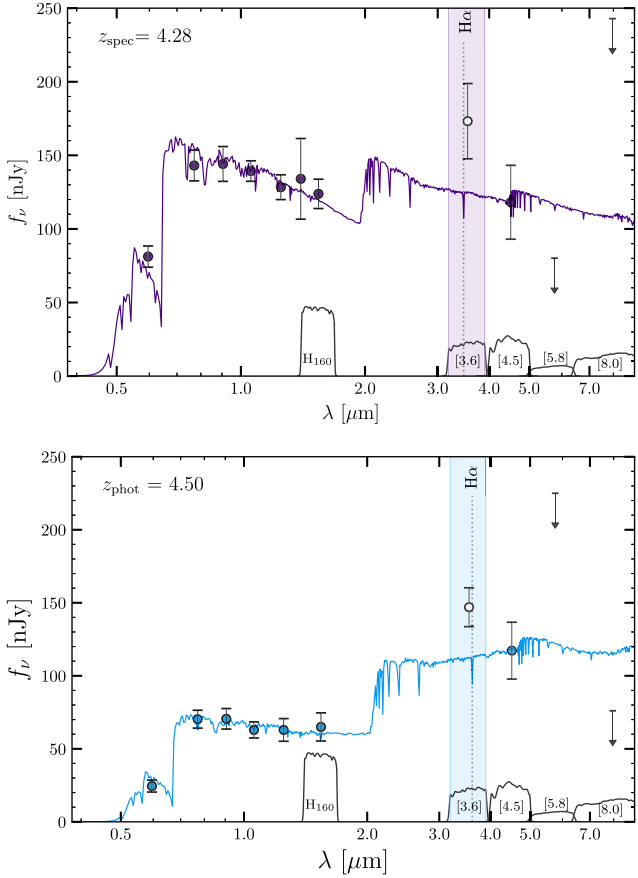

**Figure 3.** Broadband HST+IRAC photometry with their respective best-fit stellar population models for two sources in our sample. In each panel, the broadband observations are shown with filled circles, except for the IRAC [3.6] band, which is ignored in the fit to avoid the possible nebular contribution denoted by an open circle. Downward-pointing arrows denote the $2\sigma$ upper limits. The top panel is a source with known spectroscopic redshift and the bottom panel is an example from the photometric sample. The redshifted wavelength of H$\alpha$ is shown with the vertical dotted line, which falls in the range of the [3.6] IRAC band, whose wavelength range is shown by the shaded area (reference filter transmission curves are shown at the bottom of each panel). These two cases show a clear excess in the observed photometry compared to the underlying continuum of the best-fit model. This excess is primarily due to the contribution of the H$\alpha$ line to the observed flux.

**Table 1**
Correction Factors $f_q \times f_z$ per Field

| $M_{UV}$ | GSWB | GSDB | GNWB | GNDB | ERSB | XDFB |
|---|---|---|---|---|---|---|
| −22.94 | ⋯ | ⋯ | ⋯ | 0.5 | ⋯ | ⋯ |
| −22.44 | 0.5 | 0.1 | 0.1 | ⋯ | ⋯ | ⋯ |
| −21.94 | 0.1 | 4.3 | 2.1 | 9.3 | 2.9 | ⋯ |
| −21.44 | 3.6 | 3.0 | 1.9 | 3.6 | 2.8 | ⋯ |
| −20.94 | 2.3 | 2.6 | 1.9 | 2.4 | 2.7 | 3.0 |
| −20.44 | 3.1 | 2.7 | 2.9 | 2.2 | 2.9 | 3.3 |
| −19.94 | 4.0 | 4.0 | 3.8 | 3.2 | 5.0 | 3.8 |
| −19.44 | 12.4 | 7.1 | 7.3 | 5.9 | 9.2 | 4.8 |
| −18.94 | 17.6 | 55.2 | 29.5 | 97.0 | 136.0 | 5.6 |
| −18.44 | ⋯ | ⋯ | ⋯ | ⋯ | ⋯ | 16.0 |
| −17.94 | ⋯ | ⋯ | ⋯ | ⋯ | ⋯ | 15.3 |
| −17.44 | ⋯ | ⋯ | ⋯ | ⋯ | ⋯ | 67.0 |
| $M_{UV}$ | GSWV | GSDV | GNWV | GNDV | ERSV | XDFV |
| −23.36 | ⋯ | ⋯ | ⋯ | ⋯ | ⋯ | ⋯ |
| −22.86 | ⋯ | 1.0 | 1.0 | ⋯ | ⋯ | ⋯ |
| −22.36 | ⋯ | 1.0 | ⋯ | ⋯ | 1.0 | ⋯ |
| −21.86 | 8.0 | 0.4 | 0.1 | 8.0 | ⋯ | ⋯ |
| −21.36 | 6.0 | 5.0 | 21.9 | 3.7 | 5.0 | ⋯ |
| −20.86 | 9.0 | 4.6 | 25.0 | 8.6 | 6.0 | ⋯ |
| −20.36 | 5.6 | 7.4 | 16.8 | 9.3 | 7.9 | ⋯ |
| −19.86 | 9.0 | 14.2 | 15.4 | 15.3 | 8.7 | ⋯ |
| −19.36 | ⋯ | 12.2 | 34.0 | 20.6 | 44.0 | 5.0 |
| −18.86 | ⋯ | 12.7 | ⋯ | 44.5 | 10.0 | 6.6 |
| −17.86 | ⋯ | ⋯ | ⋯ | ⋯ | ⋯ | 64.0 |
| −16.86 | ⋯ | ⋯ | ⋯ | ⋯ | ⋯ | 30.0 |

**Note.** There are no correction factors when the final sample has no selected elements in that bin.

produced with important contributions from [N II] and [S II] for which we will apply a correction later. To estimates uncertainties on the H$\alpha$ fluxes, we perturb the observed photometry within error bars and adjusted the continuum level, we repeat the process 100 times, reporting the standard deviation as the $1\sigma$ uncertainty.

The flux excess measured in the [3.6] band at $z \sim 4.5$ is dominated by the contribution from H$\alpha$, [N II], and in some cases from [S II]. Adopting the ratios tabulated by Anders & Fritze-v. Alvensleben (2003) (see Table 1) for subsolar metallicity (0.2 $Z_\odot$), we estimate the fraction of the total flux that corresponds to H$\alpha$. The $z \sim 4.5$ sample includes both $B$- and $V$-band dropouts, with different redshift distributions that determine which lines contribute to the broadband flux. For the $B$-band dropouts, all three lines contribute to the [3.6] band, with H$\alpha$ accounting for 84% of the flux, consistent with the value derived by Smit et al. (2016). $V$-band dropouts are at a higher redshift and we estimate that $\sim$75% of them [S II] fall outside the [3.6] band (with an 80% certainty). For the $V$-band dropouts, then, we have only applied corrections considering the [N II] contribution to the [3.6] flux excess. In this case, H$\alpha$ accounts for 92% of the flux.

We convert the H$\alpha$ flux into H$\alpha$ luminosity using the following equation:

$$L_{H\alpha} = 4\pi \cdot D_l^2(z) \cdot F_{H\alpha}, \quad (1)$$

where $D_l^2$ is the luminosity distance, and $F_{H\alpha}$ is the flux of the line. The luminosity distance was set according to the median redshift.

The equivalent width (EW) of the H$\alpha$ line is estimated by dividing the H$\alpha$ flux derived from the procedure described above by the continuum flux density at the wavelength of the H$\alpha$ line,

$$EW_0 = \frac{F_{H\alpha}}{(1+z) f_\lambda^{cont}}, \quad (2)$$

where $F_{H\alpha}$ is the flux of the line in and $f_\lambda^{cont}$ is the flux density of the continuum at the wavelength of the redshifted H$\alpha$ emission line. The distribution of the H$\alpha$ rest-frame equivalent widths, EW$_0$(H$\alpha$), are shown in Figure 4.

$\sim$41% of the sample at $z \sim 4.5$ has observed IRAC photometry that is consistent with the synthetic best-fit SED photometry within 1$\sigma$, i.e., they either do not have emission lines or their emission lines are too weak to be detected given the photometric uncertainties (see Section 4.3). A further 6% has observed photometry at [3.6] fainter than the continuum level by more than 1$\sigma$. Finally, $\sim$53% of galaxies are detected at >1$\sigma$ in $F(H\alpha)$ for the $z \sim 4.5$ sample (Figure 4, blue histogram).

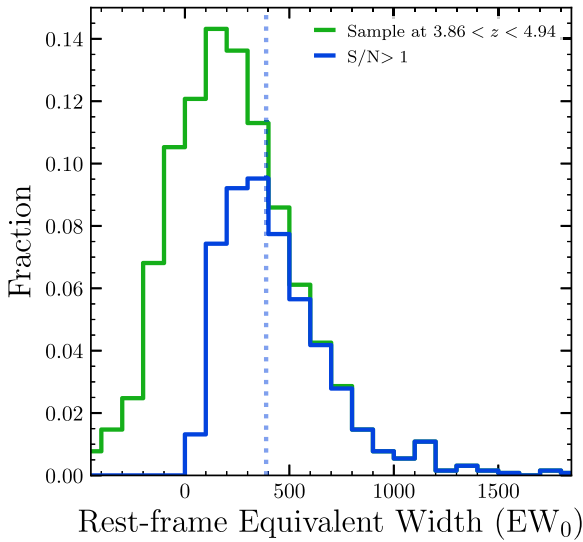

**Figure 4.** Hα rest-frame EW distribution for the sample at $z \sim 4.5$ (green histogram). 6% of the sample corresponds to negative values. Among the positive values with S/N > 1 (blue histogram), the mean value is 388 Å (shown by the dotted vertical line).

### 4.3. Detection Limit

The median uncertainty in the [3.6] photometry corresponds to $5.63 \times 10^{-22}$ [erg s$^{-1}$ Å$^{-1}$ cm$^{-2}$]. When the flux excess between the observed photometry and the estimated underlying continuum is comparable to this uncertainty, it is impossible to reliably estimate the Hα flux. We derive a detection limit based on how faint the Hα flux of an object can be and still be detected by this indirect method.

Our method is as follows: we create synthetic photometry from perturbations of the best-fit SED model of each galaxy; then we add synthetic Hα lines to the photometry with fluxes, $F_{H\alpha}$, logarithmically spaced ranging from $1.694 \times 10^{-20}$ to $1.694 \times 10^{-15}$ [erg s$^{-1}$ cm$^{-2}$]. The synthetic photometry that includes these emission lines with known fluxes is put through our pipeline to estimate Hα fluxes the same way it is done for the real sources. Finally, we can compare the input Hα flux with the estimated value. The process is repeated 50 times per galaxy SED. We find that when $F_{H\alpha} > 6.4 \times 10^{-18}$ [erg s$^{-1}$ cm$^{-2}$] we recover the Hα flux with at least a $2\sigma$ significance 60% of the time. This is almost independent of the UV luminosity of the galaxy so we have adopted this as our reference detection limit.

## 5. Hα LF

The LF is one of the most direct observables to study galaxies. It allows us to explore the evolution of their abundance and luminosity distribution over cosmic time. Here we are ultimately interested in characterizing the SFR of galaxies at $z \sim 4.5$, and for that, we make use of Hα as a tracer of star formation. This is similar to what can be done with the UV LF at this redshift (Bouwens et al. 2015) but since dust grains preferentially absorb more light at shorter wavelengths, Hα should be a more direct estimate, less sensitive to the uncertainties associated with dust extinction.

The parent sample of this study was used to derive the UV LF at $z \sim 4$ and $z \sim 5$, using the $V_{\max}$ method of Avni & Bahcall (1980) for independent samples, based on Schmidt (1968), who assigns a representative volume to each galaxy.

We adopted the comoving volumes of Bouwens et al. (2015), which already consider the incompleteness in UV detection due to the fact that faint galaxies may sometimes be lost in the image noise. Under this method, the LF is calculated according to the following equation:

$$\Phi(M) \cdot dM = \sum_i \frac{1}{V_i} = \frac{\Delta N}{\Delta V}. \quad (3)$$

To estimate the Hα LF, we start with the same $V_{\max}$ volumes associated with each galaxy to estimate the UV LF. This volume depends only on their $M_{UV}$ and redshift. Because we have made extra cuts to the sample (e.g., cuts based on IRAC S/N, quality of the SED fit), we need to correct the $V_{\max}$ volumes used. We compensate them by rescaling the volumes by the ratio between the number of sources in our sample and that in the original selection by Bouwens et al. (2015), so even after all the cuts, we recover the same UV LF from our final sample. If the cuts made to the sample were homogeneously distributed in UV luminosity, we should be able to recover the same UV LFs with our final subsample (with larger uncertainties due to using fewer objects). On the other hand, we could expect certain cuts to have an impact on the luminosity distribution. For example, fainter sources may have more uncertain redshifts and be more affected by our redshift cut.

We have verified that our final subsample reproduces the UV LFs by Bouwens et al. (2015) at $z \sim 4$ and $z \sim 5$. To do this, we have split our subsamples according to their selection as either $B$- or $V$-band dropouts. While applying the $V_{\max}$ method, we have adjusted the volumes in each magnitude bin to correct for the cuts made in the analysis (see Section 3). We assume that all missing objects in the same magnitude bin cover the same volume in every field. We do this separately for the photometric quality cuts and for the redshift cuts. As a result, the $V_{\max}$ method is modified to account for these corrections as follows:

$$\Phi(M) \cdot dM = \sum_i^{\text{Full Sample}} \frac{1}{V_i}$$
$$= \sum_i^{\text{Clean Sample}} \frac{1}{V_i} \times f_q$$
$$= \sum_i^{z-\text{selected Sample}} \frac{1}{V_i} \times f_q \times f_z, \quad (4)$$

where $f_q$ represents the correction factor for the photometric quality selection explained in Section 3.1 and $f_z$ represents the cut in redshift given by the probability that these sources have detectable Hα emission in [3.6] is greater than 80%. The final correction factors per bin of $M_{UV}$ are shown in Table 1.

Figure 5 shows UV LF that results from our final sample at $3.96 < z < 4.94$, where $B$-band dropouts and $V$-band dropouts are combined, together with the Schechter parameterizations of the UV LFs at $z \sim 4$ and $z \sim 5$ by Bouwens et al. (2015). They are in very good agreement, especially considering the slightly different redshift ranges and the reduced sample size. This same method, which reproduces the UV LFs, is the one used to estimate the Hα LFs in Section 5.

We produce the Hα LFs by binning the estimated $\log_{10} L(H\alpha)$ [erg s$^{-1}$] in bins of 0.25 and applying Equation (3) (with the corrected $V_{\max}$ volumes) to each bin. The results are shown in Figure 6 and Table 2. The detection

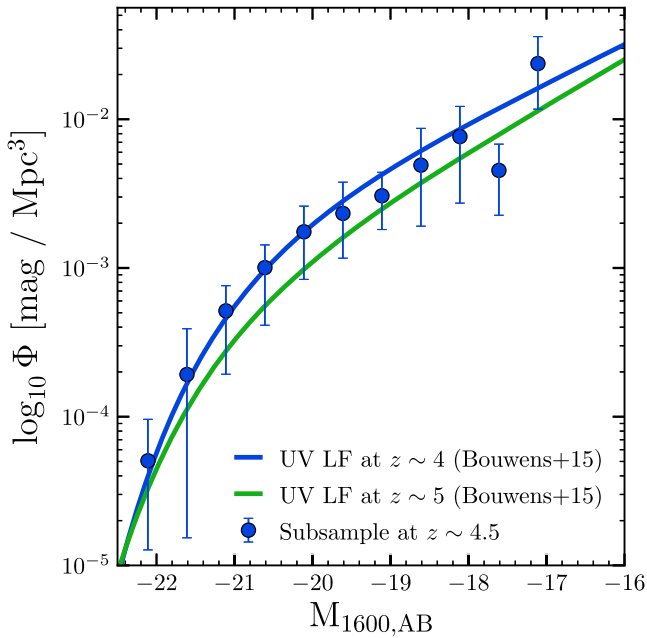

**Figure 5.** UV LF for the selected sample at the corresponding redshift range in which we can measure the Hα emission. Blue-filled circles represent the $B$- and $V$-band dropouts in the sample at $3.96 < z < 4.94$ with 1299 objects, corresponding to the UV LF recovered after adjusting correction factors for the given volume; the solid blue line represents the original UV LF at $z \sim 4$ with 5712 sources, and the solid green line represents the original curve at $z \sim 5$, which contains 1862 sources from Bouwens et al. (2015).

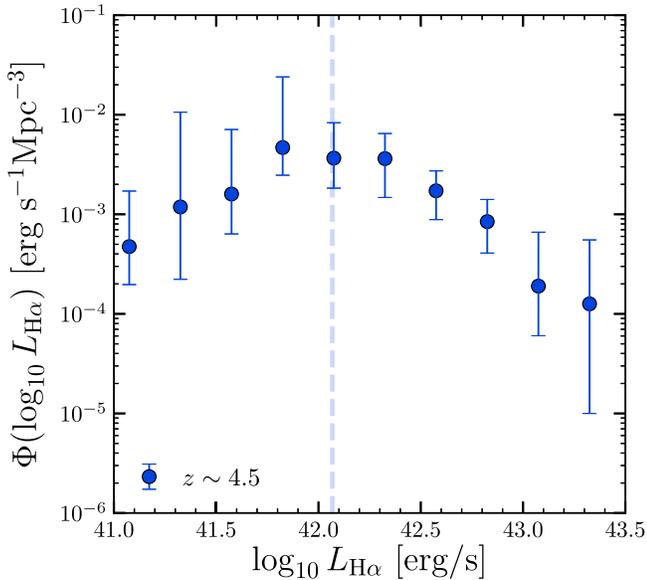

**Figure 6.** Hα LF for the sample at $z \sim 4.5$ represented by blue-filled circles. The vertical dashed line shows the detection limit derived as explained in Section 4.3. It can be noticed that below the detection limit the Hα LF decreases due to the incompleteness of the measurements in the faint end. We correct for this incompleteness in Section 5.1.

**Table 2**
Values of the Hα LF at $z \sim 4.5$

| log $L_{H\alpha}$ (erg s$^{-1}$) | $\Phi_{H\alpha}$ obs[a] ($10^{-3}$ Mpc$^{-3}$) | $\Phi_{H\alpha}$ corr[b] ($10^{-3}$ Mpc$^{-3}$) | N |
|---|---|---|---|
| | $z \sim 4.5$ | | |
| 42.075 | $4.04^{+2.47}_{-1.45}$ | $3.67^{+4.64}_{-1.84}$ | 216 |
| 42.325 | $3.24^{+1.82}_{-1.36}$ | $3.63^{+2.85}_{-2.15}$ | 235 |
| 42.575 | $1.48^{+0.83}_{-0.66}$ | $1.72^{+1.01}_{-0.83}$ | 144 |
| 42.825 | $0.34^{+0.77}_{-0.26}$ | $0.84^{+0.56}_{-0.43}$ | 83 |
| 43.075 | $0.07^{+0.53}_{-0.07}$ | $0.20^{+0.47}_{-0.13}$ | 26 |
| 43.325 | $0.06^{+0.07}_{-0.04}$ | $0.12^{+0.42}_{-0.11}$ | 9 |

**Notes.**
[a] From the observed Hα luminosity.
[b] From the dust-corrected Hα luminosity.

limits calculated in the previous section are shown as vertical dashed lines. We can see that below the detection limit there is a completeness problem, where the LF decays, inverting its faint-end slope. This incompleteness is caused by our increasing inability to measure Hα among the faintest galaxies. We attempt a correction to this incompleteness in Section 5.1.

### 5.1. Faint End of the Hα LF

Below the detection limit, the volume corrections due to incompleteness become too large and uncertain and we do not apply them. Rather, to estimate the shape of the Hα LF below this limit, we exploit the empirical relationship between $M_{UV}$ and $L_{H\alpha}$ (Figure 7, top panel). Using the $M_{UV}$–$L_{H\alpha}$ relation we compute an empirical Hα LF, in the same way that previous studies have done by bootstrap resampling (e.g., González et al. 2011; De Barros et al. 2019). This empirical Hα LF is very consistent with the obtained by the $V_{max}$ method.

Our first approach is to use the UV LF at $z \sim 4$ as a PDF to draw a sample of $M_{UV}$ values in the range of $-22.7 < M_{UV} < -16.8$. Then, we make use of the linear fit of the $M_{UV}$–$L_{H\alpha}$ relation by randomly choosing a source with a similar $M_{UV}$ (within 0.5 mag), and taking its estimated Hα flux. The result of this MC experiment is shown by the blue histogram in the bottom panel of Figure 7. As can be seen in the figure, this method yields an LF that is consistent with the one derived through the $V_{max}$ method for Hα luminosities above the detection limit. Below the detection limit, these two estimates diverge, as the MC method is not affected by the measurement incompleteness at low luminosities. We have checked that the faint-end slope estimated from this method is independent of the faint limit used when drawing samples from the UV LF.

In a second approach, we use the Schechter parameterization of the UV LF at $z \sim 4$ and combine it with a linear fit of the $M_{UV}$–$L_{H\alpha}$ relation shown by the red line in Figure 7 (top panel). This way, we can analytically derive the faint-end slope assuming a Schechter parameterization (see, e.g., González et al. 2011).

We have chosen a simple linear model to characterize the relationship between $M_{UV}$ and $\log(L(H\alpha))$ as the data is unable to constrain a more sophisticated model. We follow a Bayesian approach in which we take into account the uncertainties on both variables, assume a constant intrinsic scatter, and allow for the possibility of outliers in the sample (see Hogg et al. 2010). We model the non-detections using the Kaplan–Meier nonparametric estimator.

While we assume a constant scatter to describe this relation, recent results at $z \sim 1$ suggest an enhanced burstiness at lower masses (Atek et al. 2022), which could lead to a varying and increased scatter at the lowest luminosities. Our data does not allow us to constrain the scatter directly at the faintest

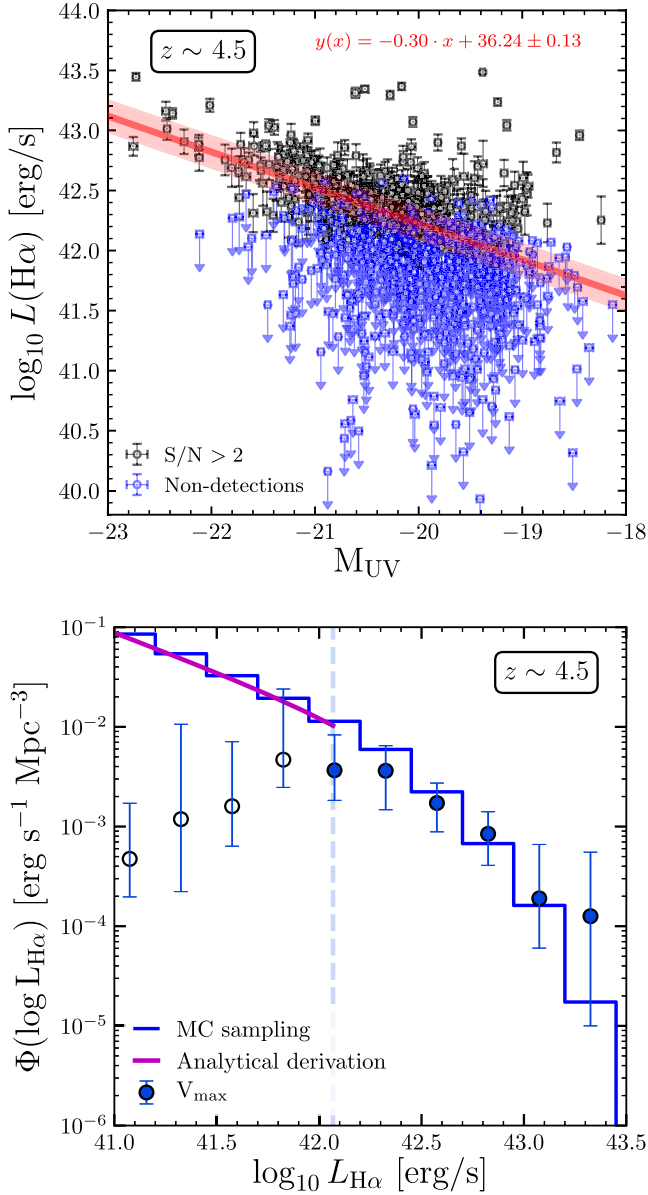

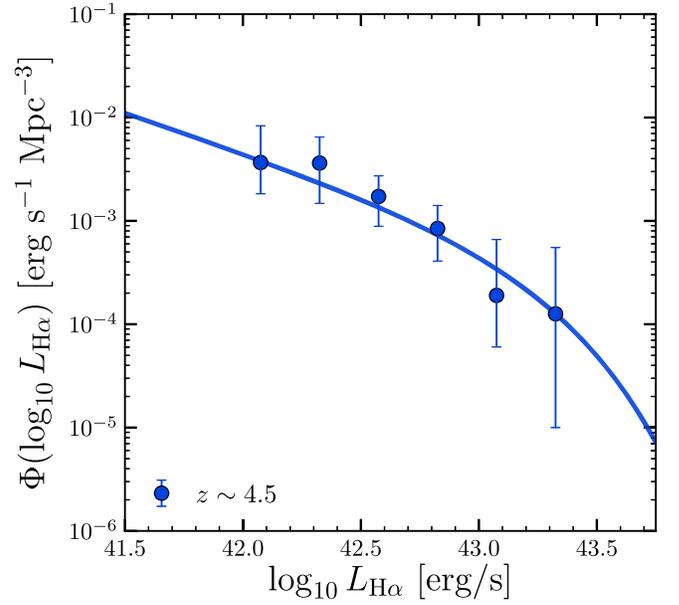

**Figure 8.** Hα LF found for the sample at $z \sim 4.5$, derived from dust-corrected luminosities. Schechter parameterization of the data with the $\alpha$ parameter fixed is done. The parameters $\Phi^*$ and $L^*$ are allowed to freely vary and the best-fit values are shown in Table 3.

**Table 3**
Schechter Parameters of the Hα LF

| $z$ | $\log_{10} L^*_{\mathrm{H}\alpha}$ (erg s$^{-1}$) | $\Phi^*_{\mathrm{H}\alpha}$ ($10^{-3}$ Mpc$^{-3}$) | $\alpha_{\mathrm{H}\alpha}$ |
|---|---|---|---|
| | Observed | | |
| 4.5 | $43.08^{+0.17}_{-0.29}$ | $0.29^{+0.69}_{-0.11}$ | $-1.83^{+0.07}_{-0.09}$ |
| | Dust corrected | | |
| 4.5 | $43.21^{+0.18}_{-0.31}$ | $0.24^{+0.76}_{-0.10}$ | $-1.76^{+0.07}_{-0.08}$ |

The result of this approach is shown in Figure 7 (bottom panel) with a magenta line. It can be seen that this approach results in an estimate of the faint-end slope that is very consistent with the MC method described above. This way, we derive the values of the faint-end slope, $\alpha$, independently of the $V_{\max}$ method (see Table 3).

### 5.2. Schechter Parameters

The Hα LF can be fitted with a Schechter function, which in the logarithmic form is

$$\Phi(L)dL = \ln(10)\phi^* \left(\frac{L}{L^*}\right)^\alpha e^{-(L/L^*)} \left(\frac{L}{L^*}\right) d\log_{10} L \quad (5)$$

with parameters $\alpha$, $\phi^*$, and $L^*$.

During the fit, the $\alpha$ parameter is fixed to the value calculated according to the process explained above in Section 5.1. To find the other best-fit parameters, an MC simulation was performed. We take 1000 random values from the simulation of the Hα LF and adjust the parameters using a least-squares method. The parameters $\phi^*$ and $L^*$ were allowed to vary freely. The Hα LF with its Schechter fit is shown in Figure 8. Each parameter with its error bar is reported in Table 3.

**Figure 7.** Top panel: the relationship between Hα luminosity and $M_{\mathrm{UV}}$. Black open circles (measurements) and blue-filled circles with arrows ($2\sigma$ upper limits) are used to fit a Bayesian linear regression. The outliers and the intrinsic scatter were also considered in the modeling. The red line corresponds to the maximum a posteriori and the shaded region represents the intrinsic scatter. The functional form including the scatter is shown at the top right of the figure. Bottom panel: the Hα LF derived from the $V_{\max}$ method (blue-filled circles), from the MC empirical sampling (blue histogram), and from the analytical derivation of the faint-end slope using the linear regression described above (magenta solid line). The detection limit is shown by the vertical dashed line.

luminosities but we have tested the possible impact of an enhanced scatter in our estimate of the faint end of the LF. To this end, we make an experiment in which we artificially increase the scatter below our detection limit by a factor of 2× compared to our best estimate. This enhanced scatter of ~0.26 dex is at the limit of what recent results show, even at the faintest luminosities (e.g., Atek et al. 2022, their Figure 9). In this experiment, we find that the new estimate of the faint-end slope is 1.03× greater than in the constant scatter model. Showing that the faint-end estimate is not very sensitive to the magnitude of the scatter, at least within the range shown by recent results.

## 6. SFR Functions

The SFR of galaxies is difficult to estimate, especially at high redshift, where the rest-frame UV is typically the only tracer readily available. UV light is strongly affected by dust extinction, which means that fairly large corrections need to be made to estimate the intrinsic UV luminosities before a conversion can be made into SFR. In this section, we estimate SFRs from the UV luminosity of our sources and we compare them to estimates derived from the H$\alpha$ luminosities derived in previous sections. Dust extinction at the wavelength of H$\alpha$ can be $\sim 3\times$ lower than in the UV (assuming, e.g., a Calzetti 1997 attenuation curve), which may make these estimates less uncertain.

### 6.1. Dust Corrections

Both, the rest-frame UV luminosity and the H$\alpha$ luminosity of a galaxy can be used to estimate its SFR. Their intrinsic values, however, are not directly observable, as dust, a key component of the interstellar medium (ISM), absorbs a significant fraction of the light emitted from the rest of the UV to the near-IR. To estimate the intrinsic UV and H$\alpha$ luminosities we must first estimate the effects of dust.

During the SED fitting procedure, which was necessary to estimate the stellar continuum at the wavelength of H$\alpha$ (see Section 4.1), dust extinction was already included in the modeling of the observed SEDs. In this case, we assumed a Calzetti et al. (2000) attenuation curve, and allowed the color excess to vary between $E(B-V) = 0$ and $0.6$. Similar to previous works we assumed that nebular lines have the same extinction as the stellar continuum, i.e., $E(B-V)_{\rm nebular} = E(B-V)_{\rm stellar}$ (e.g., Shim et al. 2011; Shivaei et al. 2015; Smit et al. 2016, but see also Calzetti 1997). We can use the results of the modeling to apply corrections to both the UV and H$\alpha$ luminosity. While this type of correction makes use of all the SED information, it is also subject to known degeneracies intrinsic to the models, in particular, the degeneracy between the age of the main stellar population and the total extinction. This means that, sometimes, similar galaxies may end up with very different dust extinction corrections if the best-fit models prefer considerably different ages.

As an alternative, we have also used the dust extinction calibration proposed by Meurer et al. (1999), which is another widely used method. Here we will focus on this method and in Appendix B we compare the SFRs derived using these two methods.

Following Meurer et al. (1999), we estimate the dust extinction at 1600 Å, in magnitudes by

$$A_{1600} = 4.43 + 1.99 \cdot \beta, \quad (6)$$

where $\beta$ is the UV continuum slope (see Section 3.2.3). To estimate the dust extinction at all other wavelengths, we use the following expression:

$$A(\lambda) = E(B-V) \cdot \kappa(\lambda), \quad (7)$$

where for $\kappa(\lambda)$, we use the Calzetti attenuation curve (Calzetti et al. 2000).

For the $z \sim 4.5$ sample, using expressions (6) and (7) the mean correction factor in the UV (1600 Å) is 1.99, and for H$\alpha$ it is 1.2. Dust extinction factors derived from SED fitting tend to be higher (Asada et al. 2021) but the results are not significantly different (see Appendix B).

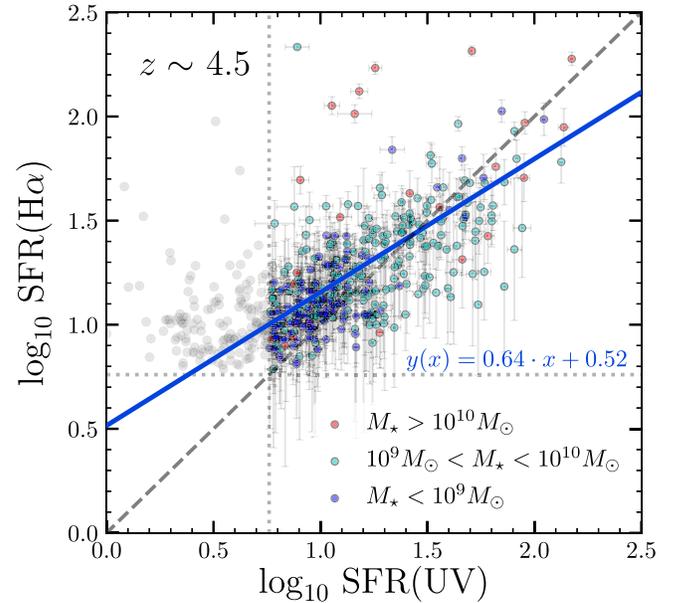

**Figure 9.** SFR derived from the H$\alpha$ luminosity in the *y*-axis vs. those derived from UV luminosity in the *x*-axis. Both luminosities were dust corrected using the Calzetti attenuation law and the IRX–$\beta$ relation. The dashed black line is the one-to-one relation, and the blue solid line shows the Bayesian linear regression (including possible outliers) and intrinsic scatter (equation shown in the bottom right). The horizontal dotted line represents the detection limit in SFR(H$\alpha$). For comparison with the SFR(UV), we have also marked the same SFR limit with a vertical dotted line.

### 6.2. SFRs

From the intrinsic UV and H$\alpha$ luminosities, we derive the SFRs. We transform the intrinsic UV luminosity into SFR following the transformation by Kennicutt (1998), scaled by a factor of 1.8 to consider a Chabrier (2003) IMF:

$$\mathrm{SFR}_{\rm UV}(M_\odot\ \mathrm{yr}^{-1}) = 0.77 \times 10^{-28} L_{\rm UV}(\mathrm{erg\ s}^{-1}\ \mathrm{Hz}^{-1}), \quad (8)$$

where $L_{\rm UV}$ is the intrinsic UV luminosity measured at 1600 Å.

Similarly, we estimate SFR from the intrinsic H$\alpha$ luminosity following Kennicutt (1998):

$$\mathrm{SFR}_{\rm H\alpha}(M_\odot\ \mathrm{yr}^{-1}) = 4.4 \times 10^{-42} L_{\rm H\alpha}(\mathrm{erg\ s}^{-1}), \quad (9)$$

where $L_{\rm H\alpha}$ is the intrinsic H$\alpha$ luminosity and the conversion also assumes a Chabrier (2003) IMF.

Figure 9 shows the comparison between the SFR derived based on the H$\alpha$ luminosity, SFR$_{\rm H\alpha}$, and the one derived using the rest-frame UV, SFR$_{\rm UV}$. While these SFR estimates are well correlated, there are clear systematic differences between them. The origin of these differences is unclear and may stem from multiple factors such as the SFHs (bursty SFHs depending on mass), metallicity trends, and variations in the attenuation curves, among others, which will be further discussed in Section 7. In the following section, we will focus on the SFRs derived using the H$\alpha$ to estimate the SFR function at $z \sim 4.5$.

### 6.3. SFR Function at $z \sim 4.5$

We derive the SFR function at $z \sim 4.5$ with the same method used to derive the H$\alpha$ LF. We bin our estimates of SFR$_{\rm H\alpha}$ in bins of $\log_{10}(\mathrm{SFR}_{\rm H\alpha}) = 0.25$ dex and adopt the same volumes used above to build the H$\alpha$ LF (see Section 5). Then, we build the SFR function with the $V_{\rm max}$ method (Equation (3)).

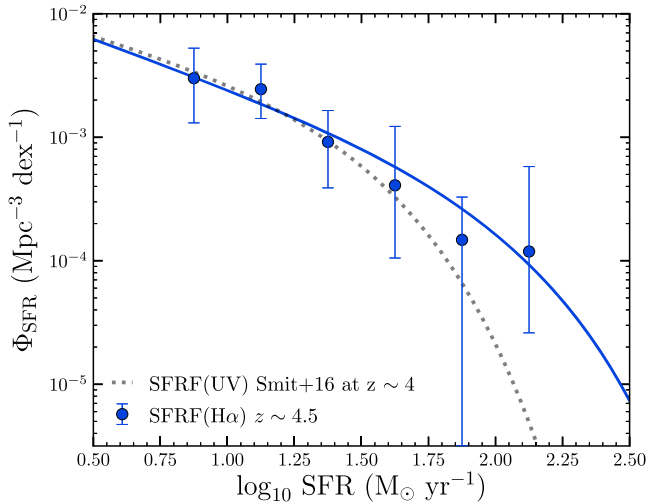

**Figure 10.** SFR function at $z \sim 4.5$ derived following the procedure described in Section 6.3. The SFR function is based on H$\alpha$ LF and we assume a Kennicutt (1998) conversion from H$\alpha$ to SFR with a correction for a Chabrier (2003) IMF. It was calculated from the stepwise dust-corrected SFR function with the analytical solution for the Schechter function (Smit et al. 2012). The Schechter function was fitted with the least-squares method considering the errors associated with each measurement and with a fixed value of the faint-end slope derived from the H$\alpha$ LF. Also, the SFR function derived from the UV at $z \sim 4$ by Smit et al. (2016) is shown as a reference.

**Table 4**
Values of the SFR Function at $z \sim 4.5$

| log SFR ($M_\odot$ yr$^{-1}$) | $\Phi_{\rm SFR}$ ($10^{-3}$ Mpc$^{-3}$) | N |
|---|---|---|
| | $z \sim 4.5$ | |
| 1.125 | $2.45^{+1.46}_{-1.03}$ | 205 |
| 1.375 | $0.92^{+0.73}_{-0.53}$ | 89 |
| 1.625 | $0.41^{+0.82}_{-0.30}$ | 44 |
| 1.875 | $0.15^{+0.18}_{-0.15}$ | 13 |
| 2.125 | $0.12^{+0.46}_{-0.09}$ | 5 |

**Table 5**
Schechter Parameters SFR Function at $z \sim 4.5$

| $\log_{10}$ SFR$^*$ ($M_\odot$ yr$^{-1}$) | $\Phi^*_{\rm SFR}$ ($10^{-3}$ Mpc$^{-3}$) | $\alpha_{\rm SFR}$ | $\rho_{\rm SFRH\alpha}$ ($M_\odot$ yr$^{-1}$ Mpc$^{-3}$) |
|---|---|---|---|
| $1.99^{+0.51}_{-0.49}$ | $0.20^{+0.60}_{-0.14}$ | $-1.76^{+0.07}_{-0.08}$ | $0.065^{+0.03}_{-0.02}$ |

Uncertainties were computed assuming a Poissonian error associated with the number of objects per bin.

To determine the completeness limit in the SFR, we use our detection limit in $L_{H\alpha}$, apply the mean dust correction factor for sources of that brightness, and convert it into SFR$_{H\alpha}$ using Equation (9). It is possible to measure SFR as low as 5.82 $M_\odot$ yr$^{-1}$, ensuring that objects are effectively detected at least with a 2$\sigma$ significance 60% of the time.

To estimate the behavior of the SFR function below the completeness limit above, we use the prescription presented by Smit et al. (2012). We perform stepwise determinations to correct H$\alpha$ luminosity in the same way that Smit et al. (2012) corrected the UV luminosity. An analytical Schechter-like approximation is used to represent the SFR functions derived from the dust-corrected H$\alpha$ LF using the relation between $A_{1600}$ and $\beta$. We do not consider a scatter for the relationship between $A_{1600}$ and $\beta$, so the slope in the faint end is obtained directly from the LF and the proper dust correction.

Similarly to what was done for the H$\alpha$ LF, we fit the Schechter function to the SFR function with parameters $\Phi^*_{\rm SFR}$, SFR$*$, and $\alpha$. We use a simple least squares method where we fix $\alpha$ to the value calculated as explained above. We allow for parameters $\Phi^*_{\rm SFR}$ and SFR$*$ to vary freely. Figure 10 shows the H$\alpha$ SFR function at $z \sim 4.5$. The values are also listed in Tables 4 and 5.

### 6.4. SFRD Evolution

The cosmic SFH of the universe shows significant scatter (Madau & Dickinson 2014) depending on the tracers used to estimate SFR. Because of observational constraints, at high redshift, most estimates rely on the rest-frame UV to estimate SFR. Here we present the SFR density based on the H$\alpha$ luminosity at $z \sim 4.5$.

To compare our H$\alpha$ based estimates at this redshift with previous UV-based estimates, we attempt to keep the same luminosity restrictions as in previous works. In particular, we integrate the SFR down to a magnitude limit $M_{\rm UV} = -17$ AB mag (Bouwens et al. 2015). In practice, we convert this magnitude limit into an SFR limit by applying a dust correction of 1.26, derived from the relation between $M_{\rm UV}$ and $\beta$ reported by Bouwens et al. (2014) for that luminosity and using Equation (8). This translates into a limit of SFR = 0.27 $M_\odot$ yr$^{-1}$. By directly integrating the best-fit Schechter functional form presented in Section 6.3 and Table 5, we find an H$\alpha$ SFRD value of 0.055 $M_\odot$ yr$^{-1}$ Mpc$^{-3}$. However, considering the asymmetric uncertainties via MC sampling we obtained a median value of 0.065 $M_\odot$ yr$^{-1}$ Mpc$^{-3}$ reported in Table 5.

Our median estimate of the H$\alpha$ SFRD at $z \sim 4.5$ is shown in Figure 11 (blue-filled point). As a comparison, we also calculated the SFRD for our sample from their UV luminosities, obtaining the blue open point shown in Figure 11. The SFRD from H$\alpha$ is 0.17 dex higher than the SFRD obtained from UV. The parameterization proposed by Madau & Dickinson (2014) of the cosmic SFH is shown by the gray curve. The values adopted by Madau & Dickinson (2014) were mainly derived from UV and IR measurements. Here we fit a new curve that only considers H$\alpha$-derived SFRDs using our new estimate at $z \sim 4.5$ in combination with those by Sobral et al. (2013) at $z \lesssim 2.5$. This new fit is shown by the red solid line and the further extrapolation is shown as a red dashed line.

### 7. Discussion

In this section, we discuss our findings regarding the evolution of the H$\alpha$ EW, and the evolution of the H$\alpha$ LF as a function of redshift. We also discuss the impact of different assumptions on our estimates of the SFR of individual galaxies and on our estimate of the SFRD over cosmic time.

#### 7.1. Evolution of H$\alpha$ EW

While there seems to be an agreement in that the specific star formation rate (sSFR = SFR/$M_{\rm stellar}$) of galaxies declines over cosmic time, it has been difficult to reconcile theoretical estimates of this decline with observations (e.g., Damen et al. 2009; Guo et al. 2011; Fumagalli et al. 2012). At the highest redshifts, the sSFR is usually estimated through SED modeling but because H$\alpha$ is a standard indicator of the SFR, the EW(H$\alpha$), can also be related to the sSFR. Studying the evolution of the EW(Hα) may be a good alternative to evaluate the evolution of the sSFR over redshift independent of SED modeling.

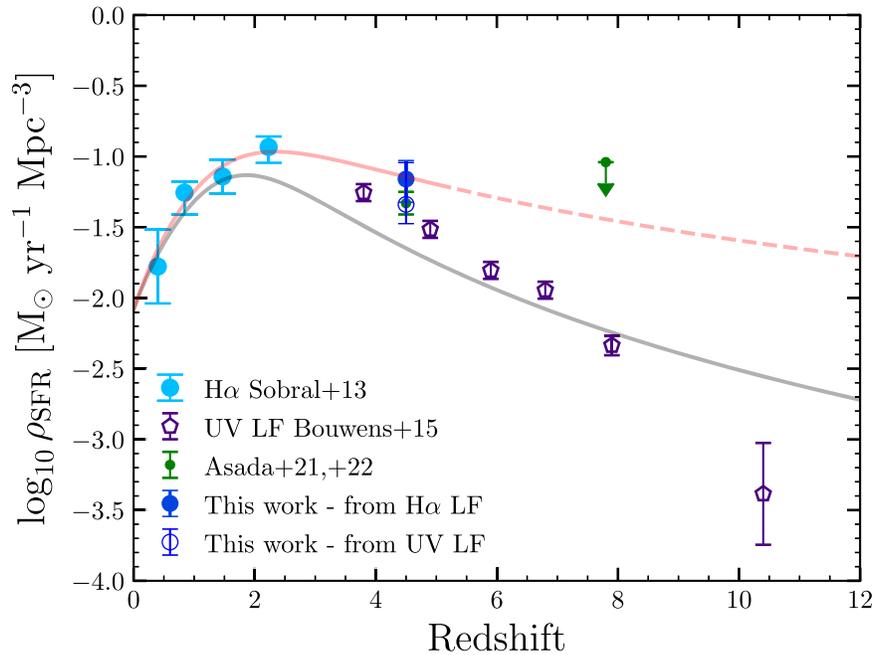

**Figure 11.** Cosmic evolution of the SFRD. Our median Hα-based estimate at $z \sim 4.5$ is shown by the solid blue circle. For comparison, we also show the median rest-UV-based estimate for our sample (open blue circle). For context, we show $z \lesssim 2.5$, Hα-based SFRD estimates by Sobral et al. (2013) ($z = 0.08, 0.4, 0.84, 1.47, 2.23$). At $z > 2.5$ there are mainly UV-based SFRDs, so here we show estimates from Bouwens et al. (2015) at $z = 3.8, 4.9, 5.9, 6.8, 7.9$ (see their Figure 18), converted into a Chabrier IMF by dividing them by 1.8. Also Asada et al. (2021) present estimations based on the rest UV to optical using SED fitting at $z \sim 4.5$, and at $z \sim 7.8$ Asada & Ohta (2022) measure the SFRD from the Hα LF. All SFRD estimates are made considering galaxies brighter than $M_{UV} = -17$, except for the case of Sobral et al. (2013) the integration is consistent with $M_{UV} = -18$. The figure also shows the functional fit for the cosmic evolution of the SFRD reported by Madau & Dickinson (2014) converted into the Chabrier IMF, which is primarily based on UV and IR estimates (gray solid line). Our fit to the Hα-based SFRD estimates is shown in red.

It is important to bear in mind that our sample is limited in brightness by the S/N > 2 cut we made in the [3.6] and [4.5] photometry. As the [4.5] band is always shallower than [3.6], once the source is detected in [4.5] we can always either measure Hα, if it is brighter than our threshold, or we can place an upper limit. Moreover, the brightness limitation in [4.5] is unaffected by emission lines; therefore, it is close to a limit in stellar mass. To make consistent comparisons with previous samples regarding the EW(Hα) we must take into account the ranges of stellar mass.

Fumagalli et al. (2012) used data from the 3D-HST survey (Brammer et al. 2012; Skelton et al. 2014) to measure the EW(Hα) from $z \sim 0$ to $z \sim 2$ and found a dramatic growth with redshift EW(Hα) $\sim (1 + z)^{1.8}$ for stellar masses in the range of $10^{10-10.5}$. Marmol-Queralto et al. (2016), however, extended the EW(Hα) measurements to $z \sim 4.5$ using the SED-fitting technique with photometry and spectroscopy of the UDS and GOODS-S fields provided by the public CANDELS and 3D-HST spectroscopic survey, and found a slower evolution EW(Hα) $\sim (1 + z)^{1.0}$ for star-forming galaxies with stellar masses $\simeq 10^{10} M_\odot$. Their estimates at $z \sim 4.5$ are significantly lower than those reported by Shim et al. (2011), which they argue is due to the improved quality of their $K_s$-band data. Other studies have also found estimates below those reported by Shim et al. (2011). In particular, the Stark et al. (2013) estimate at $z \sim 5$ is very consistent with that by Marmol-Queralto et al. (2016) but they are both slightly lower than the estimates presented by Smit et al. (2016). This seems to be explained by the lower stellar masses ($\sim 0.9$ dex) in the latter works. Controlling by mass seems to produce more consistent results.

Our work extends the estimates of the EW(Hα) to $z \sim 4.5$ using the latest Spitzer/IRAC data. The stellar mass of our sample is in the range of $10^{8.5-10.5} M_\odot$. Figure 12 shows our median value of EW(Hα) for the full sample as a function of redshift with the blue-filled point. By restricting the stellar mass range we draw the median value of EW(Hα) in the lowest stellar masses ($10^{8.5-9.5} M_\odot$) with an open point and with a filled magenta point for the highest stellar masses ($10^{10.0-10.5} M_\odot$). We find an EW(Hα) value that is very consistent with those of Smit et al. (2016) and Marmol-Queralto et al. (2016) at $z \sim 4.5$. This may be expected given the similar mass ranges considered in these studies ($\simeq 10^{8.5-9.5} M_\odot$). Our measurements, however, do not allow us to distinguish between the evolution proposed by Fumagalli et al. (2012) and the one proposed by Marmol-Queralto et al. (2016), as they are formally consistent with both. Our measurements suggest an evolution at $z > 4$ that is in between the slower evolution estimated by Marmol-Queralto et al. (2016) and the faster one by Fumagalli et al. (2012).

### 7.2. Evolution of Hα LF

Sobral et al. (2013) studied the evolution of the Hα LF using deep and wide narrow-band filters from the High-redshift (Z) Emission Line Survey at $z = 0.4, 0.84, 1.47,$ and 2.23. Here we can extend the study of the Hα LF evolution to $z \sim 4.5$.

Figure 13 compares our best-fit Schechter function at $z \sim 4.5$ with the $z = 2.23$ LF by Sobral et al. (2013). While the $z = 2.23$ Hα LF is above the others at the faint end ($< 10^{43} L_{H\alpha}$), it seems that bright Hα (strong emitters and likely higher SFRs) are more common at the highest redshifts. In terms of the evolution of the best-fit Schechter parameters (Figure 14), we find that the normalization factor, $\Phi^*$, continues to decrease toward higher redshift, consistent with its behavior at $z > 1$. In fact, shows a decreasing behavior from $z = 0.84$–4.5 by a factor of $\sim$14. In the case of $L^*$, the evolution shows a steady increase with redshift up to $z \sim 4.5$ as shown in the middle panel of Figure 14. Finally, the evolution of the faint-end slope, $\alpha_{H\alpha}$ is shown in the bottom panel of Figure 14. In this case, our value does not come from a Schechter fit but a power-law fit based on our MC method (see Section 5.1). $\alpha_{H\alpha}$ shows a fairly constant value as a function of redshift, although the value at $z \sim 4.5$ may be slightly lower than all $z < 3$ estimates, implying steeper H$\alpha$ LFs at the highest redshifts.

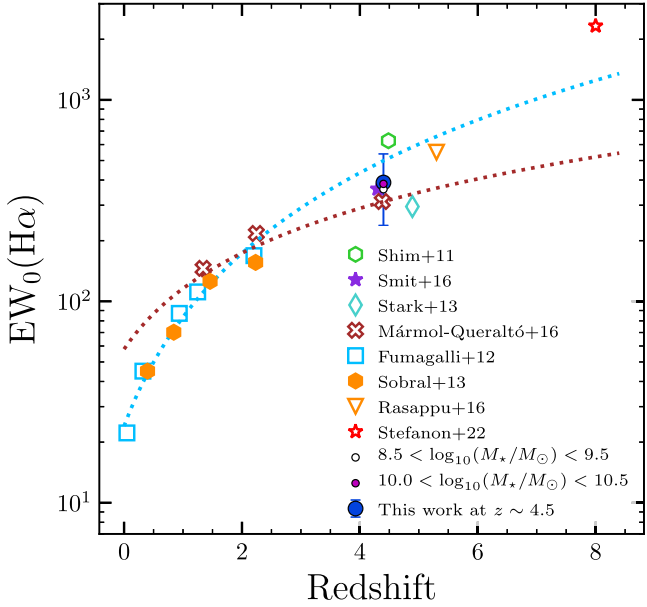

**Figure 12.** Evolution of EW(H$\alpha$) with redshift. The blue-filled point represents our median EW(H$\alpha$) measurements for the full sample at $z \sim 4.5$. Additionally, the small black open circle and the magenta-filled point show the median EW(H$\alpha$) controlled by stellar mass in the ranges $10^{8.5-9.5}$ and $10^{10-10.5}$ $M_\odot$, respectively. Results from other H$\alpha$ studies are shown for comparison (Shim et al. 2011; Fumagalli et al. 2012; Sobral et al. 2013; Stark et al. 2013; Smit et al. 2016; Marmol-Queralto et al. 2016; Rasappu et al. 2016; Stefanon et al. 2022). The dotted lines show the evolution of this quantity proposed by Fumagalli et al. (2012), for the stellar masses range given by $10^{10-10.5}$ and S/N > 3 data of their work (light blue curve), and by Marmol-Queralto et al. (2016, red curve).

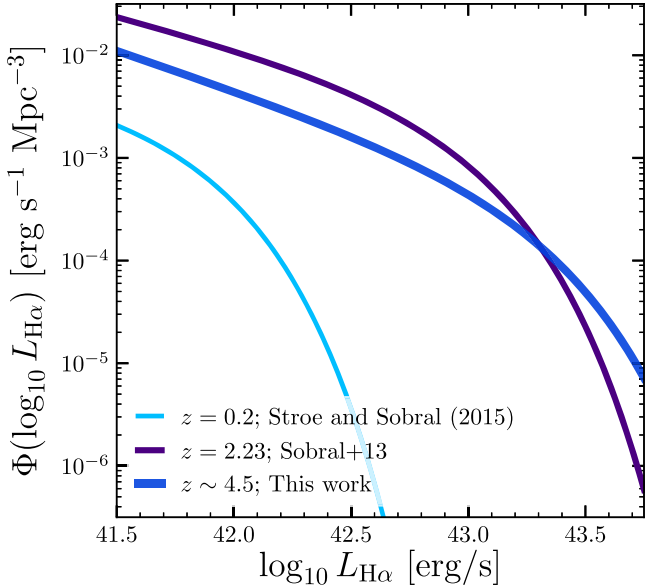

**Figure 13.** Our H$\alpha$ LF (Schechter fits) at $z \sim 4.5$ compared to the H$\alpha$ LF by Sobral et al. (2013) at $z \sim 2.2$. The H$\alpha$ LF at $z = 0.2$ presented by Stroe & Sobral (2015) is shown as a representative reference of the local universe. There is a clear evolution of the normalization factor of the LF, also the *knee* of the function changes according to redshift, and the faint-end slope does not show a clear evolution.

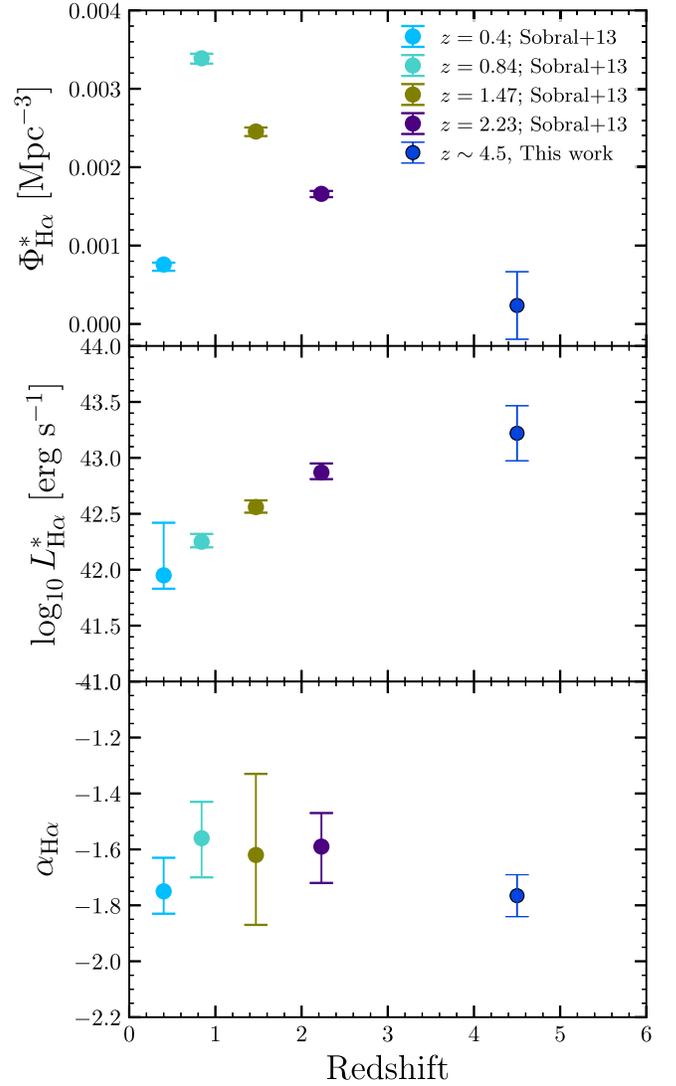

**Figure 14.** The evolution of the Schechter best-fit parameters for the H$\alpha$ LF since $z \sim 4.5$. Top panel: the evolution of $\Phi^*$ shows a decrease with redshift from $z \sim 1$. Middle panel: the evolution of $L^*_{H\alpha}$ shows a consistent increase up to $z \sim 4.5$. Bottom panel: the faint-end slope shows a fairly flat behavior as a function of redshift.

### 7.3. Differences between SFR(H$\alpha$) and SFR(UV)

In this work, we have derived SFRs both from UV luminosities and from H$\alpha$ fluxes (see Section 6.2). H$\alpha$-based SFRs are sensitive to changes in the recent SFH and can change in relatively short timescales of 10 Myr. UV SFRs, on the other hand, depend on the luminosity of slightly lower mass stars and so they are more representative of the SFR over the past $\sim$100 Myr. If the SFHs are changing over timescales comparable to ∼10 Myr, we can expect these two SFR estimates to differ, and in fact, we may learn about the variability of the SFHs of galaxies based on these differences.

Previous studies have evaluated the relationship between UV-derived SFRs and Hα-derived SFRs and found differences between both indicators. Atek et al. (2022), for example, compared the SFR$_{UV}$ with the SFR$_{H\alpha}$ in low-mass ($<10^9$ $M_\odot$) galaxies at $0.7 < z < 1.5$. They find that these low-mass galaxies have an excess in SFR$_{H\alpha}$ compared to SFR$_{UV}$ and that they also tend to have higher EW(Hα). They also find that the excess becomes larger toward lower masses. This could suggest that the SFHs of lower mass galaxies are more *bursty*. Similarly, Faisst et al. (2019) analyzed a sample of galaxies at $4 < z < 5$ and based on statistical arguments concluded that a galaxy at this redshift has experienced on average 1–4 major bursts.

Smit et al. (2016) also report excess in Hα-derived SFRs compared with UV-derived SFRs at $3.8 < z < 5.0$ but they argue that bursty SFHs may not be the best explanation for their sample. They find that a simple model that adds bursts to the SFH of galaxies may reproduce the excess in SFR$_{H\alpha}$ but it would also create a fraction of sources with low sSFR(Hα) that are not seen in their data.

Figure 9 shows, for our galaxies, the comparison between SFR(Hα) and SFR(UV). The horizontal line shows our estimated completeness limit (see Section 5.1). Galaxies with SFR below this limit could be detected in UV but would not be detected in Hα unless they have a significant excess in SFR (Hα) compared to SFR(UV). This would bias the comparison and make the SFR(Hα) excess look particularly strong at the lowest values of SFR(UV). This can be seen as denoted by the gray points in Figure 9. Even ignoring these grayed-out points below the detection limit, however, there is an interesting trend in the SFR(Hα)/SFR(UV) ratio. At the highest SFRs (log(SFR (UV) >1.5), SFR(Hα) is consistently lower than SFR(UV), indicating that their SFRs may have been declining in the last 10 Myr. On the other end, for the lowest SFRs that we can measure, SFR(Hα) is enhanced compared to SFR(UV). This may indicate that they are more likely to be experiencing a burst at the time of observation, or that their SFHs are more typically rising. Because SFR is correlated with $M_{stellar}$, this could mean that lower mass systems tend to have more bursty SFHs, as suggested by Atek et al. (2022), but the rising SFHs are also consistent. This change between (on average) rising SFHs at low mass to declining SFHs at the highest masses is consistent with the behavior shown in the semi-analytic simulation by Tacchella et al. (2018). Our data, however, also show significant scatter on this trend, especially among the most massive galaxies ($M_{stellar} > 10^{10}$ $M_\odot$). For intermediate and low-mass galaxies the behavior seems more consistent.

Unfortunately, it is not so straightforward to derive conclusions regarding the SFH of galaxies based on the differences between SFR(Hα) and SFR(UV), as there are other factors that can affect these SFR estimators. For example, Ly et al. (2016) suggest that at one-fifth of solar metallicity, the conversion to SFRs from Hα is ∼0.2 dex lower than the Kennicutt (1998) relation that we used. We find that this factor is not enough to balance SFR$_{UV}$ and SFR$_{H\alpha}$.

Another important factor affecting the comparison could be the differential dust attenuation factor between stellar continuum and nebular emission, which has been reported to be $A_{nebular} = 0.4 A_{continuum}$ (e.g., Calzetti et al. 2000). Other recent studies at $z \sim 2$, however, do not find a significant difference between the dust attenuation for stellar continuum and for nebular emission (Shivaei et al. 2015), which may depend on the metallicity (Shivaei et al. 2020). Nevertheless, for this effect to explain our observed trend in SFR(Hα)/SFR(UV) ratio without the necessity to invoke changes in the SFHs, they would have to also be dependent on SFR or stellar mass. For example, differential dust attenuation between stellar continuum and nebular emission would have to become larger as a function of increasing SFR. This would bring the slope of the SFR(Hα)–SFR(UV) relation closer to 1 but it would still mean that SFR(Hα) is higher than SFR(UV) at all SFRs.

Finally, there is the possibility that the attenuation curve is different from that of Calzetti et al. (2000) assumed here. For example, previous studies have explored the possibility of a steeper attenuation curve in high-$z$ galaxies (e.g., Shivaei et al. 2015, 2020; Smit et al. 2016). Smit et al. (2016) concluded that if high-$z$, UV-selected galaxies had an SMC-type dust law, the Hα fluxes observed would be too high to be explained without a stronger source of ionizing radiation. We find that testing an SMC attenuation law in our galaxies (Gordon et al. 2003), does not resolve the tension between SFR(Hα) and SFR(UV) either, unless the steepness of the dust attenuation law is also dependent on SFR or stellar mass. This is also true with even steeper laws.

### 7.4. CSFRD History

By integrating the SFR functions presented in Section 6.3 we can estimate the SFRD of the universe at $z \sim 4.5$ based on SFR (Hα). The limit adopted for the integral should be consistent with previous UV-based estimates at $z > 3$ ($M_{UV} < -17$, see Section 6.4). As can be seen in Figure 11, Hα based estimates are systematically higher than the UV-based estimates at all redshifts. The estimates presented at $z < 3$ are derived from the Hα LFs presented by Sobral et al. (2013) and using the same conversions used in this work. At $z \sim 4.5$ Asada et al. (2021) estimate the SFRD from SED fitting showing a good agreement with our estimation from the UV LF. However, our estimate from the Hα LF with our selected sample is higher than the value they report.

Because we have worked with a subsample of the original $B$- and $V$-band dropouts from the work of Bouwens et al. (2015), we have tested if our sample is consistent with their SFRD trends. We find a very good agreement at $z \sim 4.5$.

As discussed in the previous section, individual Hα based SFRs are lower than UV-based estimates at high values of SFR but the opposite is true at the lowest SFRs. Due to the steepness of the SFR function, the low SFR sources dominate the numbers, which could explain why the integral results in SFRD (Hα) are higher than the SFRD(UV) when integrated into the ranges used in Figure 11.

Similarly to Madau & Dickinson (2014), we have fit all the SFRD(Hα) values from $z \sim 0.4$ to $z \sim 4.5$ and found a best fit:

$$\psi(z) = 0.015 \cdot \frac{(1+z)^a}{1 + [(1+z)/2.9]^b} \; M_\odot \; \text{yr}^{-1} \; \text{Mpc}^{-3}. \quad (10)$$

This functional form is essentially the same as the one presented by Madau & Dickinson (2014) but $a$ and $b$ are equal to 3.0 and 4.7 instead of 2.7 and 5.6. A very similar result is found if we used a different dust extinction correction based on the SED fitting instead (see Appendix B). Asada & Ohta (2022) place an upper limit at $z \sim 7.8$ derived from the H$\alpha$ luminosity, and although it was not considered in the fit, our extrapolation is consistent with the fact that the SFRD from the H$\alpha$ luminosity is higher than the obtained from the UV.

## 8. Summary

In this paper, we investigate the H$\alpha$ emission in LBG-selected galaxies at $z \sim 4.5$. We have estimated H$\alpha$ from the excess flux in the photometry at [3.6] $\mu$m compared to the best-fit SEDs, which was obtained with CIGALE (Boquien et al. 2019). We have used the deepest Spitzer/IRAC imaging available at [3.6] and [4.5] over the GOODS fields from the GREATS program (Stefanon et al. 2021). We have used our estimates of the H$\alpha$ flux in galaxies at $z \sim 4.5$ to estimate rest-frame EW(H$\alpha$) and the H$\alpha$ LF. We have also applied dust corrections based on the Meurer et al. (1999) relation and the dust attenuation curve by Calzetti et al. (2000). Combined with standard Kennicutt relations we have made estimates of the SFR(H$\alpha$) and the SFR LF. We have integrated the SFR function to calculate the SFRD at $z > 4$ as estimated from H$\alpha$. Our main findings and conclusions are the following:

1. The rest-frame EW of H$\alpha$ evolves with redshift, having higher values at higher redshift. Our estimate of the EW(H$\alpha$) at $z \sim 4.5$ is in between those presented by Fumagalli et al. (2012) and the ones by Marmol-Queralto et al. (2016). Given the uncertainties, our estimate is consistent with both previous estimates. We also find a tentative correlation such that lower mass galaxies have higher EW(H$\alpha$).
2. The H$\alpha$ LF evolves with redshift showing a decreasing normalization, $\Phi^*$, from $z = 0.84$–$4.5$ by a factor of $\sim 14$. Meanwhile, $L^*$ increases with redshift from the local universe to high-$z$, and while our bright end is not very well constrained, it shows an excess of bright H$\alpha$ emitters compared to the LF at $z \sim 2$. The faint-end slope $\alpha$ shows no significant evidence of any changes.
3. We have estimated both the SFR$_{H\alpha}$ and the SFR$_{UV}$ using standard conversion factors. We find that galaxies with $\log(\mathrm{SFR(UV)}) < 1.5$ tend to have a higher SFR(H$\alpha$) compared to SFR(UV). The opposite is also true for the higher SFRs. While this may be a sign of increased burstiness among the lower SFR galaxies (also the lower stellar masses), this could also show a difference in the typical SFH of galaxies as a function of mass, with lower mass galaxies having more rising SFHs, and higher mass galaxies having SFHs that are declining at later times.
4. The CSFRD derived from H$\alpha$ luminosity tends to be systematically elevated compared to the CSFRD estimated from the rest-frame UV. This is true at $z < 2$ based on the results by Sobral et al. (2013) compared with the curve proposed by Madau & Dickinson (2014) (mainly based on UV and IR measurements), but the difference becomes more significant at higher redshift.

While this work makes use of the deepest mid-IR imaging available, it is still limited in S/N, only allowing us to estimate H$\alpha$ for galaxies with SFR(H$\alpha$) $\gtrsim 6$ $M_\odot$ yr$^{-1}$. Future spectroscopic observations with JWST will significantly improve the measurements of H$\alpha$ in fainter systems and will also extend their study to higher redshifts.

V.G. gratefully acknowledges support by the ANID BASAL projects ACE210002 and FB210003. M.S. and R.J.B. acknowledge support from TOP grant TOP1.16.057. M.S. acknowledges support from the CIDEGENT/2021/059 grant, from project PID2019-109592GB-I00/AEI/10.13039/501100011033 from the Spanish Ministerio de Ciencia e Innovación - Agencia Estatal de Investigación. This study forms part of the Astrophysics and High Energy Physics programme and was supported by MCIN with funding from European Union NextGenerationEU (PRTR-C17.I1) and by Generalitat Valenciana under the project No. ASFAE/2022/025. P.A.O. acknowledges support from the Swiss National Science Foundation through the SNSF Professorship grant 190079 "Galaxy Build-up at Cosmic Dawn." The Cosmic Dawn Center (DAWN) is funded by the Danish National Research Foundation under grant No. 140. We also acknowledge the support of NASA grants HSTAR-13252, HST-GO-13872, HST-GO-13792, and NWO grant 600.065.140.11N211 (vrij competitie). R.S. acknowledges an STFC Ernest Rutherford Fellowship (ST/S004831/1). G.D.I. acknowledges support for GREATS under RSA No. 1525754. This paper utilizes observations obtained with the NASA/ESA Hubble Space Telescope, retrieved from the Mikulski Archive for Space Telescopes (MAST) at the Space Telescope Science Institute (STScI). STScI is operated by the Association of Universities for Research in Astronomy, Inc. under NASA contract NAS 5-26555. This work is based [in part] on observations made with the Spitzer Space Telescope, which was operated by the Jet Propulsion Laboratory, California Institute of Technology under a contract with NASA. Support for this work was provided by NASA through an award issued by JPL/Caltech.

## Appendix A
## Alternative H$\alpha$ Measurements

We propose a new method to derive the H$\alpha$ flux from broadband photometry that is independent of the SED modeling. The main idea is to build a predictive model to estimate the rest-frame continuum level at 6563 Å.

To estimate the continuum level we use, as a reference, galaxies whose redshift is in the range of $3.0 \leqslant z \leqslant 3.7$, for which the IRAC 3.6 $\mu$m measures the stellar continuum flux without a significant contribution from nebular emission: the *clean sample*. We select 279 sources with photometric redshift in this range (with >80% probability). Based on these galaxies we train a model to predict the continuum level at 6563 Å. The rest-frame optical SED is fairly flat and the redshift range is very close to the sample at $3.86 < z < 4.94$ so this is an appropriate reference. In essence, this model interpolates the stellar continuum at 6563Å using as reference a sample of galaxies at similar redshift without relying on SED fitting.

We then apply the model to the sample of galaxies for which [3.6] does include H$\alpha$ contribution and we measure the observed excess flux at [3.6] compared with the model prediction. The following model is proposed:

$$f_1(f_2, f_{\mathrm{UV}}, \beta) = A \cdot \frac{f_2^2}{f_{\mathrm{UV}}} + B \cdot \beta \cdot f_2 + C \cdot f_{\mathrm{UV}} \cdot f_2 + D \cdot f_2, \quad (\mathrm{A1})$$

where $f_1$ is the flux density at 3.6 $\mu$m, which we want to predict, $f_2$ is the observed flux density at 4.5 $\mu$m, $f_{\mathrm{UV}}$ is the rest-frame flux at 1600 Å, and $\beta$ is the UV-continuum slope from the fit $f_\lambda \propto \lambda^\beta$. A–D are the free parameters of the model.

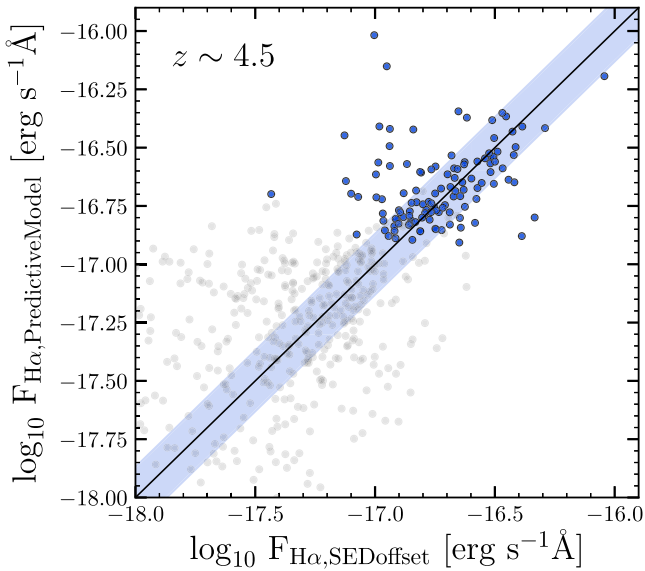

**Figure 15.** Hα fluxes derived from the predictive model vs. the fiducial estimates used throughout, based on the offset between the observed photometry and the best-fit SED. Blue-filled circles have $2\sigma$ significance, and the gray points have less than $2\sigma$ significance, but they still follow the same trend as the circles. Both methods are consistent on average, with a scatter of 0.13 dex (see text) that is comparable to the typical uncertainties. This shows that the Hα flux estimates used throughout do not depend strongly on the details of the SED fitting procedure.

We train the predictive model with 80% of the clean sample and then we test the value of the flux at [3.6] with the remaining 20%. To estimate the error of the model we iterate this procedure 100 times, perturbing the parameters within their associated uncertainties.

We use the flux excess at [3.6] and repeat the same procedure to calculate the Hα as described in Section 4, including the same corrections to account for [N II] and [S II] in the [3.6] band.

Figure 15 shows the comparison between the fiducial estimates used in the paper (based on the stellar continuum estimated with CIGALE, horizontal axis), and the method described above (vertical axis). The black line is the identity for reference. Only sources detected with $>2\sigma$ significance are shown as blue-filled circles, and the rest of the gray points have $<2\sigma$. We find no systematic offsets between the two estimates but there is considerable scatter. Focusing only on sources with $>2\sigma$ Hα fluxes, the scatter of the relation is 0.13 dex (median absolute deviation), with 14.3% of outliers ($>2\times$ MAD). This scatter is comparable to the mean uncertainty of our Hα estimates. These results suggest that the Hα flux estimates used throughout do not depend strongly on the details of the SED fitting procedure and that the uncertainties are not significantly underestimated.

### A.1. Modeling Nebular Emission with CIGALE

While for our fiducial estimates we have run CIGALE to model the stellar continuum only, excluding the bands that may include significant nebular emission line contribution, the code also has the option to include nebular emission in the SED fitting. We have run a separate model that includes nebular emission lines and fit the SEDs, including all bands. From here, we can read out directly the $F_{H\alpha}$ from the best fits.

Figure 16 compares our fiducial Hα flux estimates used throughout with those produced by CIGALE using models that include nebular emission. Overall, the comparison shows significant scatter and a systematic offset such that CIGALE estimates produce slightly higher Hα fluxes. Interestingly, the bigger the discrepancy, the higher the $\chi^2$ of the best-fit model (as shown by the color coding of the circles in Figure 16). This probably suggests that most of the discrepancy is a result of the difficulty of the models to reproduce the Hα EWs suggested by the broadband photometry.

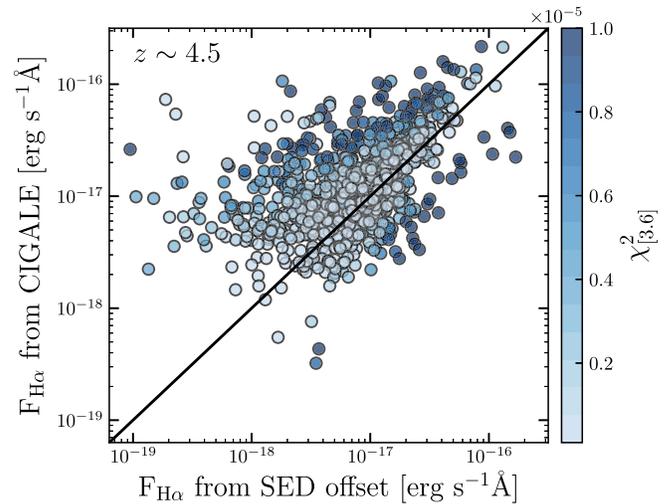

**Figure 16.** Comparison between our fiducial Hα fluxes and the ones derived from SED modeling with CIGALE, including nebular emission. Color code represents the value of $\chi^2$ from the corresponding band where Hα lies. This quantity reveals how well the best-fit model reproduces actual photometry in each band.

## Appendix B
## CIGALE Dust Corrections

Our SFR estimates, in particular our SFRD estimates, depend on the corrections done to account for the effects of dust. Our fiducial result is done using the UV slope $\beta$ to estimate $A_{1600}$ using the Meurer et al. (1999) relation and the Calzetti et al. (2000) attenuation curve. Independently, we also estimate the dust correction while doing the SED modeling with CIGALE (see Section 4.1). Here we explore the consistency between both dust correction estimates.

Figure 17 shows that the comparison between the Hα luminosities was corrected using the two different extinction values. CIGALE estimates of the dust corrections are slightly higher, resulting in intrinsic Hα luminosities that are on average $1.3\times$ higher. Propagating these small differences in dust correction to the results of our SFRD estimates we obtain a consistent estimation that is $1.27\times$ higher as shown Figure 18. The final estimate is not very sensitive to our choice between these two estimates of dust correction.

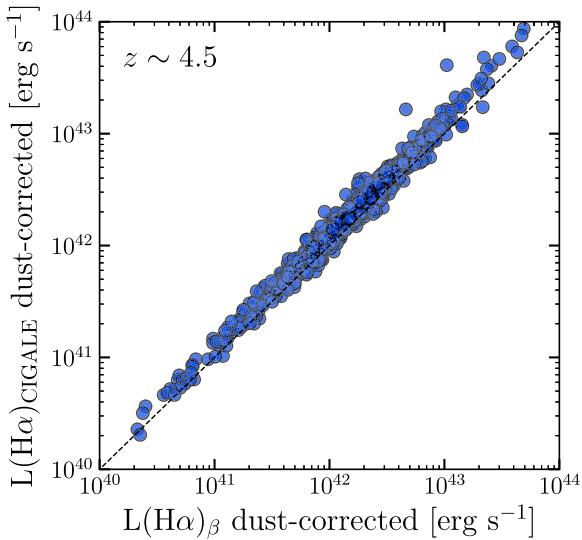

**Figure 17.** Hα luminosities dust corrected using CIGALE best-fit model compared to the fiducial Hα luminosities used throughout which are based on the (Meurer et al. 1999) relation. CIGALE-based corrections result in Hα luminosities on average 1.3× higher.

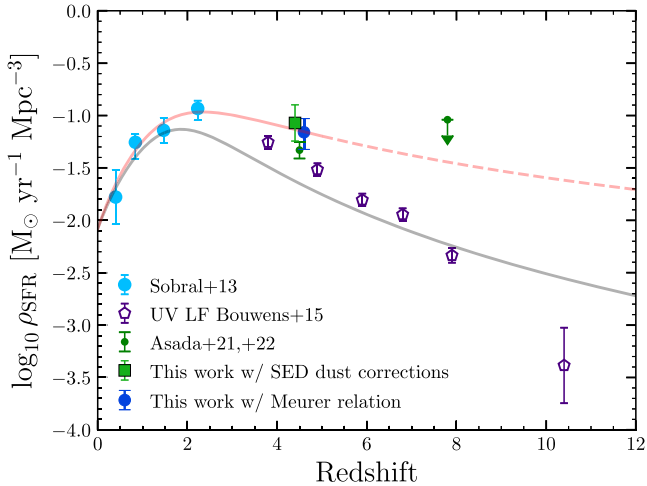

**Figure 18.** Cosmic evolution of the SFRD with the luminosities corrected by dust from the extinction derived by CIGALE instead of the fiducial result presented in the main text, based on the Meurer et al. (1999) relation. This is equivalent to Figure 11 but with the alternative dust corrections. The final result does not depend strongly on the choice between these two dust corrections.

### ORCID iDs

Victoria Bollo ● https://orcid.org/0000-0002-9842-296X
Valentino González ● https://orcid.org/0000-0002-3120-0510
Mauro Stefanon ● https://orcid.org/0000-0001-7768-5309
Pascal A. Oesch ● https://orcid.org/0000-0001-5851-6649
Rychard J. Bouwens ● https://orcid.org/0000-0002-4989-2471
Renske Smit ● https://orcid.org/0000-0001-8034-7802
Garth D. Illingworth ● https://orcid.org/0000-0002-8096-2837